\documentclass{aa}
\usepackage{txfonts}
\usepackage{graphicx,lscape}
\usepackage{amsmath}
\usepackage{natbib}
\usepackage{amssymb}
\bibpunct{(}{)}{;}{a}{}{,} % to follow the A&A style
\begin{document}

\title{Evolution of surface CNO abundances in massive stars}
\titlerunning{CNO abundances}

\author{Andr\'e Maeder\inst{1}, Norbert Przybilla\inst{2,3}, Mar\'ia-Fernanda Nieva\inst{2,3}, Cyril Georgy\inst{4},\\ Georges Meynet\inst{1}, Sylvia Ekstr\"{o}m\inst{1}, Patrick Eggenberger\inst{1}}
\authorrunning{Maeder et al.}

\institute{Geneva Observatory, Geneva University, CH--1290 Sauverny, Switzerland\\\email{andre.maeder@unige.ch} \and Institute for Astro- and Particle Physics, University of Innsbruck, Technikerstr.~25/8, A-6020 Innsbruck, Austria \and Dr.~Karl Remeis-Observatory\,\&\,ECAP, University Erlangen-Nuremberg, Sternwartstr.\,7, D-96049 Bamberg, Germany \and Astrophysics group, EPSAM, Keele University, Lennard-Jones Labs, Keele, ST5 5BG, UK\\
}

\date{Received ; accepted }

\abstract
{}
{The nitrogen to carbon (N/C) and nitrogen to oxygen (N/O) ratios are the most sensitive quantities to mixing in stellar interiors of intermediate and massive stars. We further investigate the theoretical properties of these ratios as well as put in context recent observational results obtained by the VLT-FLAMES Survey of massive stars in the Galaxy and the Magellanic Clouds.}
{We consider analytical relations and numerical models of stellar evolution as well as our own stellar atmosphere models, and we critically re-investigate observed spectra.} 
{On the theoretical side, the N/C vs~N/O plot shows little dependence on the initial stellar masses, rotation velocities, and nature of the mixing processes up to relative enrichment of N/O by a factor of about four, thus this plot constitutes an ideal quality test for observational results. The comparison between the FLAMES Survey and theoretical values shows overall agreement, despite the observational scatter of the published results. The existence of some mixing of CNO products is clearly confirmed, however the accuracy of the data is not sufficient for allowing a test of the significant differences between different models of rotating stars and the Geneva models. We discuss reasons (for the most part due to observational bias) why part of the observational data points should not be considered for this comparison. When these observational data points are not considered, the scatter is reduced. Finally, the N/C vs~N/O plot potentially offers a powerful way for discriminating blue supergiants before the red supergiant stage from those after it. Also, red supergiants of similar low velocities may exhibit different N enrichments, depending on their initial rotation during the main-sequence phase.}
{}


\maketitle
 
%%%%%%%%%%%%%%%%%%%%%%%%%%%%%%%%%%%%%%%%%%%%%%%%%%%
 \section{Introduction}
%%%%%%%%%%%%%%%%%%%%%%%%%%%%%%%%%%%%%%%%%%%%%%%%%%%

Models for the evolution of massive stars are of utmost importance for wide fields of astrophysics. Among many others, they provide the basis for the interpretation of stellar populations in stellar clusters and galaxies \citep{MConti94}, define the structure of supernova progenitors and nucleosynthesis yields \citep{Bresolin08}. They are the starting point for predicting the properties of the first generation of stars \citep{BrommL04} and for explaining exotic phenomena such as neutron stars, stellar black holes, and $\gamma$-ray bursts of the long-duration soft-spectrum type \citep{Woosley12}.

Differences in the massive star populations in various galactic environments (like blue-to-red supergiant ratios or Wolf-Rayet to O star numbers) have long since indicated that the standard theory of massive star evolution \citep[\textit{e.g.},][]{ChMa86}, which describes the stars as mass-losing non-rotating bodies parameterised by mass and chemical composition (metallicity), is not complete. More recently, progress in astrophysical observation, in particular from high-resolution spectroscopy and from asteroseismology, has shown more deviations from standard theory, \textit{e.g.},~in form of nitrogen enrichments of the surface layers of massive stars near the main sequence (MS).

Over the past two decades, a growing body of evidence has been assembled indicating that rotation \citep{MeMa00,HeLa00,MeMa05}, binarity \citep{Wellsteinetal01}, and possibly magnetic fields \citep{MaMe05,Hegeretal05,BrakingMEM11} are as important factors for massive star evolution as mass-loss. Past developments are reviewed by \citet{MaMe00}, \citet{maederlivre09}, and \citet{MaMe12} for single stars, and by \citet{Vanbeverenetal98} and \citet{Langer12} in consideration of additional effects of massive binary star evolution. \citet{MaPa13} recently presented a quantitative comparison of currently available model grids, aimed at constraining the uncertainties on the predicted evolutionary paths from the theoretical side.

The importance and sources of internal mixing is a major concern of most recent models of stellar evolution, since mixing affects all of the model outputs. The models make detailed predictions of the surface properties (stellar parameters and chemical abundances) of massive stars that need to be compared to detailed observations for verification of the assumptions made in the models. So far, the most ambitious effort in this respect was the VLT-FLAMES Survey of massive stars in the Galaxy and the Magellanic Clouds \citep[in brief, the `FLAMES Survey' in the following,][]{Evansetal05,Evansetal06}. The quantitative analysis of a large number of early B-type stars covering a wide range of rotational velocities gave rise to challenges in the concept of rotational mixing \citep{Hunteretal08b,Hunteretal09}. In particular, the existence of highly nitrogen-enriched slow rotators \citep[see also][]{Moreletal06,NiPr12,RiveroGonzalezetal12} and nitrogen-normal fast rotators (together composing $\sim$40\% of the sample stars) was found to contradict current model predictions.

\citet{Maederetal09} pointed out that the behaviour of the excess $N/H$ ratio, which is the primary evidence for this criticism, is a multivariate function 
\begin{eqnarray*}
\Delta \log (N/H) = f (M, \tau_\text{evol}, v\sin i,
\text{multiplicity}, Z)\,,
\end{eqnarray*}
\noindent where $M$ is the mass of a star, $\tau_\text{evol}$ its age, $V\sin i$ the (projected) rotational velocity, and $Z$ the metallicity. In order to find a correlation of a multivariate function with a parameter ($V\sin i$ in the case of the FLAMES Survey), it is necessary to limit the range of the other parameters involved as much as possible. \citet{Maederetal09} argued that the concept of rotational mixing is in fact supported, when the range of parameters is limited.

However, the enrichment of nitrogen is only one aspect of rotational mixing. With nitrogen, carbon and oxygen also have to follow tight correlations as participants in the CNO-bicycle, in addition to helium, which is the burning product \citep{Przy2011}. The most sensitive parameters to mixing are the nitrogen to carbon (N/C) and nitrogen to oxygen (N/O) ratios. In this respect, the N/C~vs~N/O~plot of observed abundances is essential. This shall be studied further here.

We first investigate the properties of the N/C vs~N/O diagram, both by analytical (Sect.~\ref{nuclearconstraints}) and numerical approaches (Sect.~\ref{numericalmodels}). In particular, we examine its sensitivity to the initial stellar masses, to the metallicity, and to the strength of mixing. We compare the recent determinations of CNO abundances for OB stars within the FLAMES Survey to the theoretical predictions and we try to analyse the discrepancies (Sect.~\ref{comparisonobservation}). In Sect.~\ref{assessment}, we perform our own line-formation computations for a $\sim$10\% fraction of early B-type stars from the FLAMES Survey to re-assess the published results by \citet{Hunteretal09}. Finally, the conclusions from our investigations are summarised in Sect.~\ref{conclusions}.

%%%%%%%%%%%%%%%%%%%%%%%%%%%%%%%%%%%%%%%%%%%%%%%%%%%
\section{Nuclear constraints on the N/C vs~N/O plots\label{nuclearconstraints}}
%%%%%%%%%%%%%%%%%%%%%%%%%%%%%%%%%%%%%%%%%%%%%%%%%%%

First, we examine the changes of the N/C vs N/O ratios predicted by the CNO cycle analytically. Two different simplifying hypotheses can be made. In the case of the most massive stars, the carbon $^{12}$C immediately reaches equilibrium by the CN cycle and is turned to $^{14}$N \citep[see for example][]{maederlivre09}. The oxygen $^{16}$O is slowly destroyed during the MS to produce nitrogen $^{14}$N. Thus, we take the number C of carbon atoms as a constant, and note by N and O the numbers of nitrogen and oxygen atoms. We consider only the most abundant isotopes of each of these elements. We have 
 \begin{eqnarray}
 \text{d}(\text{N}/\text{C})= \text{dN}/\text{C} \; ,\\
 \text{dO}= -\text{dN} \; .
 \end{eqnarray}
One has to be careful about treating mass fractions and number ratios correctly. For example, with mass fractions, the above relations would be written:
 \begin{eqnarray}
 \text{d}(X_\text{N}/X_\text{C})= \text{d}X_\text{N}/X_\text{C} \; ,\\
 \text{d}X_\text{O}= -\frac{8}{7} \text{d}X_\text{N} \; .
 \end{eqnarray}
 \noindent The constant number of carbon atoms implies the constancy of the mass fraction $X_\text{C}$. Each time a nitrogen atom is destroyed an oxygen atom is created, but their mass fractions are related as indicated, since turning $^{14}$N to $^{16}$O requires the addition of two nucleons, at the expense of the H-content. 

\begin{figure}[t!]
\begin{center}
\includegraphics[width=.99\linewidth]{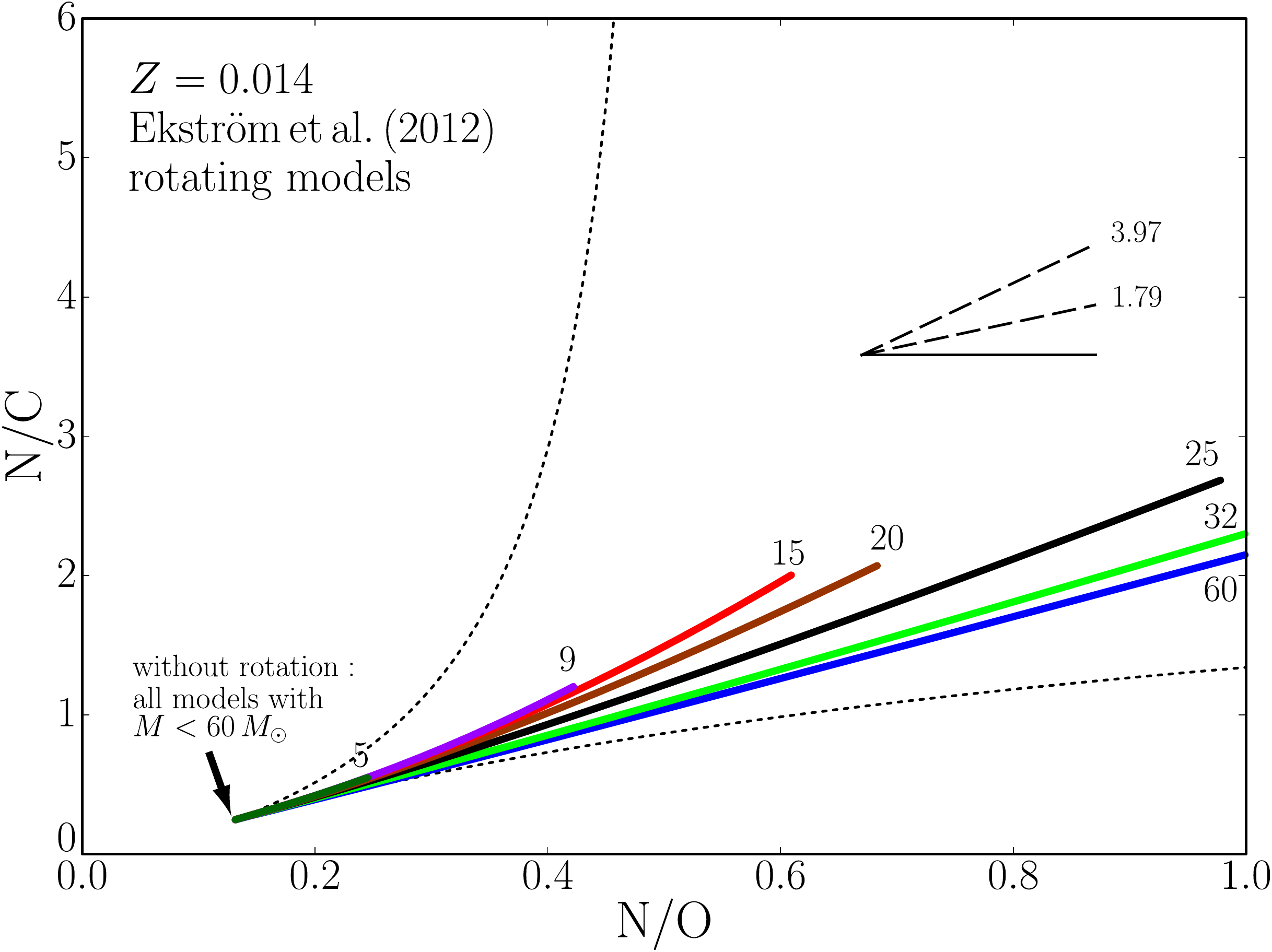}
\caption{The N/C vs~N/O abundances (in numbers) during the MS phase of models of rotating stars with $Z=0.014$ and initial masses in the range of 5 to 60\,$M_{\sun}$ by \citet{Grille2012}. The initial rotation velocities correspond to 40\% of the critical velocity. The slopes (\ref{slopenum}) obtained from the analytical approximations are indicated. The two dotted lines in black indicate the limiting solutions, (\ref{solanalOconst}) for the lower line, and (\ref{solanalCconst}) for the upper line. They are correct near the initial abundances, but then rapidly and largely deviate from the numerical models.}
\label{modelesNCNO}
\end{center}
\end{figure}

\begin{figure}[t!]
\begin{center}
\includegraphics[width=.99\linewidth]{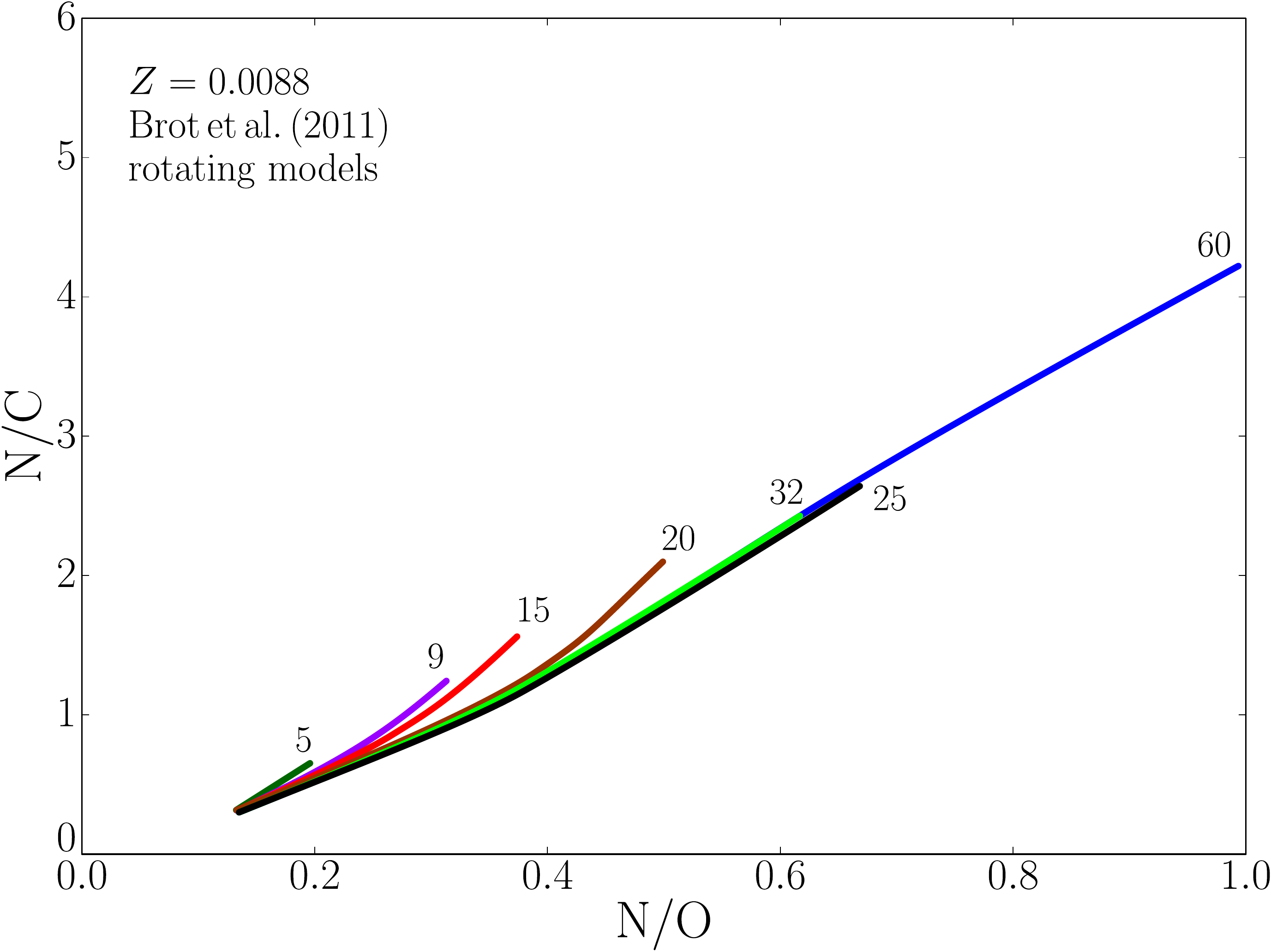}
\caption{The N/C vs~N/O abundances (in numbers) during the MS phase of models of rotating stars with $Z=0.0088$ for the Galaxy and initial masses in the range of 5 to 60\,$M_{\sun}$ by \citet{Brott2011}. The curves shown correspond to initial rotation velocities of 40\% of the critical velocities for comparison with Fig. \ref{modelesNCNO}.}
\label{BrottMW}
\end{center}
\end{figure}

In terms of numbers, we have
\begin{eqnarray}
\text{d}\left(\frac{\text{N}}{\text{O}}\right) =\frac{\text{dN}}{\text{O}}- \frac{\text{N}}{\text{O}^2} \,\text{dO} 
=\frac{\text{dN}}{\text{O}}\left(1+ \frac{\text{N}}{\text{O}} 
\right)\; ,
\end{eqnarray}
\begin{eqnarray}
\frac{\text{d}\left(\frac{\text{N}}{\text{C}}\right)}{\text{d}\left(\frac{\text{N}}{\text{O}}\right)}
=\frac {\text{N}/\text{C}}{\text{N}/\text{O}} \, \frac{1}{\left(1+ \frac{\text{N}}{\text{O}}\right)} \; .
\label{highm}
\end{eqnarray}
\noindent This means that the slope in a plot N/C vs~N/O initially only depends on nuclear physics and initial CNO ratios\footnote{The reader may wonder why appear here the initial values of N/C and N/O, and not the values obtained after CN equilibrium. Actually, it can be shown that when the dilution factor is near 1, then the values of N/C and N/O appearing in the equations are actually initial values. The dilution factor is defined as the fraction of the mass with initial composition divided by the total mass which is mixed. This total mass is made from the mass with initial composition (initially in the envelope) and the mass with CN equilibrium composition (initially in the core).}. In such a plot for massive stars, the first term on the right produces a linear relation, while the second one flattens the slope as the CNO cycle proceeds. We shall see this flattening of the curve for the most massive stars in Figs.~\ref{modelesNCNO}, \ref{BrottMW}, \ref{MODELMC}, \ref{BrottLMC}, \ref{MODELSMC}, and \ref{BrottSMC} obtained from numerical models. There, the slopes for the most massive stars are lower than for intermediate masses, because of the more rapid transformation of oxygen into nitrogen in more massive stars.\\

Another simplifying assumption applies to the lower mass stars experiencing the CNO cycle. There, one may assume that the CNO cycle starts by converting $^{12}$C to $^{14}$N, keeping the $^{16}$O abundance about constant. This implies in numbers \citep{Przy2011},
\begin{eqnarray}
\text{dC}= - \text{dN} \quad \text{and} \\ \text{d}(\text{N}/\text{O}) = \text{dN}/\text{O}\;.
 \end{eqnarray}
 Thus, we have
 \begin{eqnarray}
\text{d}\left(\frac{\text{N}}{\text{C}}\right)= %\frac{dX_N}{X_C}- \frac{X_N}{X^2_C} \frac{dX_C}{dX_N} dX_N =
\frac{\text{dN}}{\text{C}}\left(1+ \frac{\text{N}}{\text{C}} \right)\;,
\end{eqnarray}
\noindent leading to the ratio
\begin{eqnarray}
\frac{d\left(\frac{N}{C}\right)}{d\left(\frac{N}{O}\right)}=\frac{\text{N}/\text{C}}{\text{N}/\text{O}}
\left(1+ \frac{\text{N}}{\text{C}}\right)\;.
\label{lowm}
\end{eqnarray}
\noindent There, the relation deviates upwards from a linear curve as more advanced products of the CNO cycle become visible. Figures in Sect. \ref{numericalmodels} show well this effect for the lower stellar masses (5-15\,$M_{\sun}$) in the range considered.\\

 With rotation included \citep{Grille2012}, the new models predict N-enrichment down to 2\,$M_{\sun}$. For a moderate enrichment corresponding, for example, to N/O=0.3 and N/C=0.7 in intermediate and massive stars (see Fig.~\ref{modelesNCNO}), we find from Eqs.~(\ref{highm}) and (\ref{lowm}), respectively, the following slopes:

\begin{eqnarray}
\frac{\text{d}\left(\frac{\text{N}}{\text{C}}\right)}{\text{d}\left(\frac{\text{N}}{\text{O}}\right)}
= 1.79,\quad \text{and} \; \; 3.97 \;.
\label{slopenum}
\end{eqnarray}
\noindent These local average slopes, based on our two simplifying assumptions, are illustrated in Fig.~\ref{modelesNCNO}. The slope for the lower mass stars is steeper than that of the higher mass stars, in agreement with the analytical predictions.

%: fig 1

We may also note that Eqs.~(\ref{highm}) and (\ref{lowm}) can be integrated analytically. We call $y$=N/C and $x$=N/O. Eq.~(\ref{highm}) with C constant becomes 
\begin{eqnarray}
\frac{\text{d}y}{\text{d}x}= \frac{y}{x} \, \frac{1}{1+ \,x} \;,
\end{eqnarray}
\noindent which admits a solution of the form
\begin{eqnarray}
y=\frac{k_{1} \, x}{1+ \, x} \; , % \quad \text{with} \; k_1 =\frac{y_{\text{ini}}}{x_{\text{ini}}} \left( 1+ \frac{8}{7} \, x_{\text{ini}} \right)\; .
\end{eqnarray}
\noindent where $k_1$ is a constant defined by initial conditions. This is
\begin{eqnarray}
\frac{\text{N}}{\text{C}}\, = \, k_1	 \, \frac{\frac{\text{N}}{\text{O}}}{1+\frac{\text{N}}{\text{O}}}
\quad \text{with} \; \; \;
k_1\, = \, \frac{ \left(\frac{\text{N}}{\text{C}}\right)_\text{ini}\left[1+\left(\frac{\text{N}}{\text{O}}\right)_\text{ini}\right]
}{\left(\frac{\text{N}}{\text{O}}\right)_\text{ini}} \; .
\label{solanalCconst}
\end{eqnarray}
\noindent With the same notations for $x$ and $y$, Eq.~(\ref{lowm}) with O constant becomes
\begin{eqnarray}
\frac{\text{d}y}{\text{d}x}= \frac{y}{x} \left( 1+ \,y \right) \; .
\end{eqnarray}
\noindent This equation has the following solution:
\begin{eqnarray}
y=\frac{x}{k_2 - \, x} \; ,
\end{eqnarray}
\noindent where $k_2$ is another constant determined by the initial conditions. This is
 \begin{eqnarray}
 \frac{\text{N}}{\text{C}}\, = \frac{\frac{\text{N}}{\text{O}}} {k_2-\frac{\text{N}}{\text{O}}} \; ,
 \quad \text{with} \; \; \;
 k_2\, = \, \frac{\left(\frac{\text{N}}{\text{O}}\right)_\text{ini}}{\left(\frac{\text{N}}{\text{C}}\right)_\text{ini}} +\left(\frac{\text{N}}{\text{O}}\right)_\text{ini} \; .
 \label{solanalOconst}
 \end{eqnarray}
 
The two solutions (\ref{solanalCconst}) and (\ref{solanalOconst}) are plotted in Fig.~\ref{modelesNCNO}, which also presents the results of the numerical models discussed in the next Section. We see that these two solutions are tangent and encompass the model results close to the initial abundances, but they rapidly and largely deviate upwards and downwards from the numerical solutions when CNO-processed elements appear. This is not a surprise in view of the simplifying assumptions made. The analytical solutions are only valid for small departures from the initial abundances. The numerical models show intermediate values between the two limiting cases, as normally expected. We notice in this context that average slopes (\ref{highm}) and (\ref{lowm}) over a limited interval are likely better simplifying assumptions, as illustrated by Fig.~\ref{modelesNCNO}. 

From the point of view of stellar physics, we recall that the N/C and N/O ratios do not always change simultaneously and in the same way. The branching ratio between the ON loops and the CN cycle is about $10^{-3}$ at a typical temperature $T=2.5 \cdot 10^{7}$ K. Thus, the conversion of C to N occurs at a lower temperature $T$ than that of O to N. Thus, the main effect in massive stars is that the conversion of C to N occurs much faster than that of O to N. The reactions also occur differently at different levels in the stars, but this has generally little consequence for massive stars on the MS, since efficient nuclear reactions occur in a large convective core. This may be different in later evolutionary phases. Thus, depending on the strength of the mixing processes at different levels in the stars and at different ages, the relations between the observed surface N/C and N/O ratios may be different. This explains why these relations are not always the same and may show differences as evolution proceeds.

%: fig 2

%%%%%%%%%%%%%%%%%%%%%%%%%%%%%%%%%%%%%%%%%%%%%%%%%%%
\section{Models for OB stars in the Galaxy, LMC, and SMC\label{numericalmodels}}
%%%%%%%%%%%%%%%%%%%%%%%%%%%%%%%%%%%%%%%%%%%%%%%%%%%

%============================================================================
\subsection{Numerical models for solar abundances} \label{solarm}
%============================================================================

Recent grids of the Geneva models \citep[][hereafter E12]{Grille2012}\defcitealias{Grille2012}{E12} consider both non-rotating and rotating stars in the range of mass 0.8 to 120\,$M_{\sun}$. The rotating models have an initial velocity equal to 40\% of the critical equatorial velocity. Three different sets of models have been made or are in progress with initial metallicities $Z=0.014$ \citepalias{Grille2012}, $Z=0.006$ (Eggenberger et al. in prep.) and $Z=0.002$ \citep{GeorgyZ002}. Solar or scaled solar elemental abundances are adopted with the initial N/C and N/O ratios for solar abundances of 0.248 and 0.131 (by number), respectively. 

For $Z=0.014$, the initial mass fractions of the main CNO elements are $X(^{12}\text{C})=2.283\cdot 10^{-3}$, $X(^{14}\text{N})=6.588\cdot 10^{-4}$, and $X(^{16}\text{O})=5.718\cdot 10^{-3}$ \citep{Asplundetal05,Asplundetal09}. Figure~\ref{modelesNCNO} shows the relations N/C vs N/O during the MS for models from 5 to 60\,$M_{\sun}$. This means that the present results apply to clusters with ages between 4 and 110\,Myr. For non-rotating stars, there is no enrichment in this mass interval, except at the end of the MS phase of the 60\,$M_{\sun}$ model (there the change of surface composition is due to mass loss).

%: fig 3
\begin{figure}[t]
\begin{center}
\includegraphics[width=.98\linewidth]{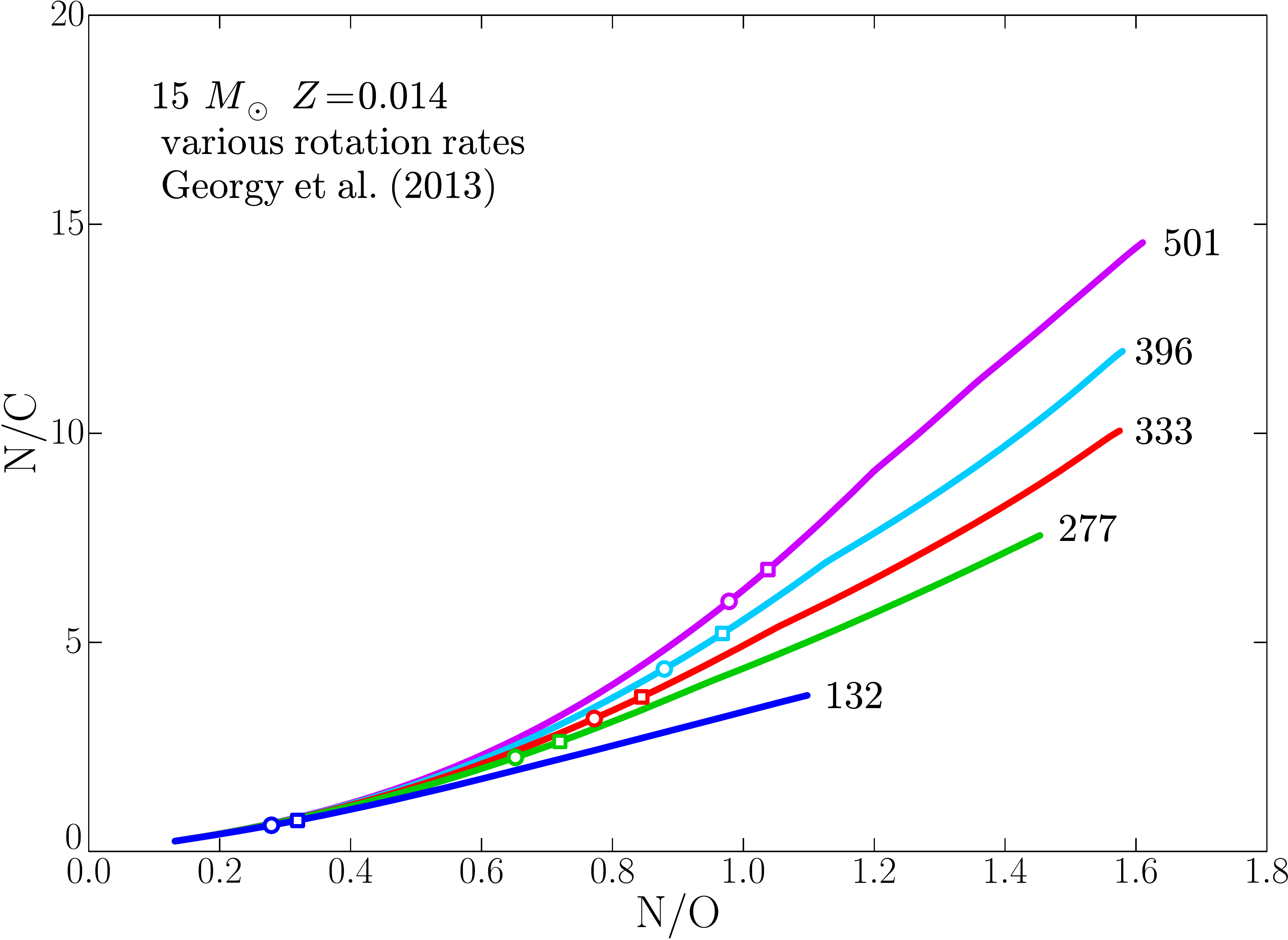}
\caption{The N/C vs~N/O abundances (in numbers) for a 15\,$M_{\sun}$ model with $Z=0.014$ \citep{GrilleBe} with various initial rotation velocities (in km\,s$^{-1}$). Open circles indicate the end of the MS phase and squares mark the beginning of the red supergiant phase.}
\label{M15014rot}
\end{center}
\end{figure}

The rotating stellar models \citepalias{Grille2012}, with an initial velocity of 40\% of the critical one\footnote{The reader should be aware that the definition of $V_\text{crit}$ used by \citet{Brott2011} and the definition used by the Geneva group are different and not exactly comparable.}, correspond rather well to the average observed velocities at each mass. As illustrated by Fig.~\ref{modelesNCNO} the relations are very close to linear for the whole MS phase. At 5\,$M_{\sun}$, the mixing is small and N/O reaches only 0.26 at the end of the MS phase. In agreement with relations (\ref{slopenum}), the less massive stars have the steepest slopes, while the relation is flatter for massive stars. The range of slopes span by the models during the MS phase is rather narrow. It is a bit smaller than the range given by the analytical approximation (\ref{slopenum}). This is understandable, since the analytical approximations apply to two extreme cases and the numerical models are between these extremes.

Figure~\ref{BrottMW} is identical as Fig. \ref{modelesNCNO} but shows the models by \citet[hereafter B11]{Brott2011}\defcitealias{Brott2011}{B11}, in some cases interpolations have been made from the grid of these authors. The metallicity of $Z=0.0088$ adopted by these authors is lower than the metallicity adopted in the Geneva models and based on \citet{Przybilla2008} and \citet{NiPr12}, however the adopted CNO ratios are similar in both sets of models. The initial slope is the same as in the Geneva models, but it is a bit steeper in models by \citetalias{Brott2011} away from the zero age MS. This may result from the different physics of mixing in the two sets of models. The main differences are: 1) the much larger overshoot parameter adopted by \citetalias{Brott2011}; 2) the Geneva models adopt an advecto-diffusion treatment for the meridional circulation, while \citetalias{Brott2011} use a diffusion scheme; and 3) the models by \citetalias{Brott2011} include magnetic field while the Geneva models do not. The large initial overshooting (since the pressure scale height $H_P$ is larger initially) favours a rapid surface increase of N/C by the CN cycle, while the contribution of the ON cycle becomes relatively more important in the course of evolution.

The extension of the tracks in Figs.~\ref{modelesNCNO} and \ref{BrottMW} is a measure of the importance of mixing. We see that the N/C enrichments are about the same in both sets of models. However, the enrichments in N/O are larger, by at least 50\%, in the Geneva models for the models with an average (rather low) velocity equal to 40\% of the critical velocity. The effects are different for higher velocities, as shown below.

%: fig 4
\begin{figure}[t]
\begin{center}
\includegraphics[width=.98\linewidth]{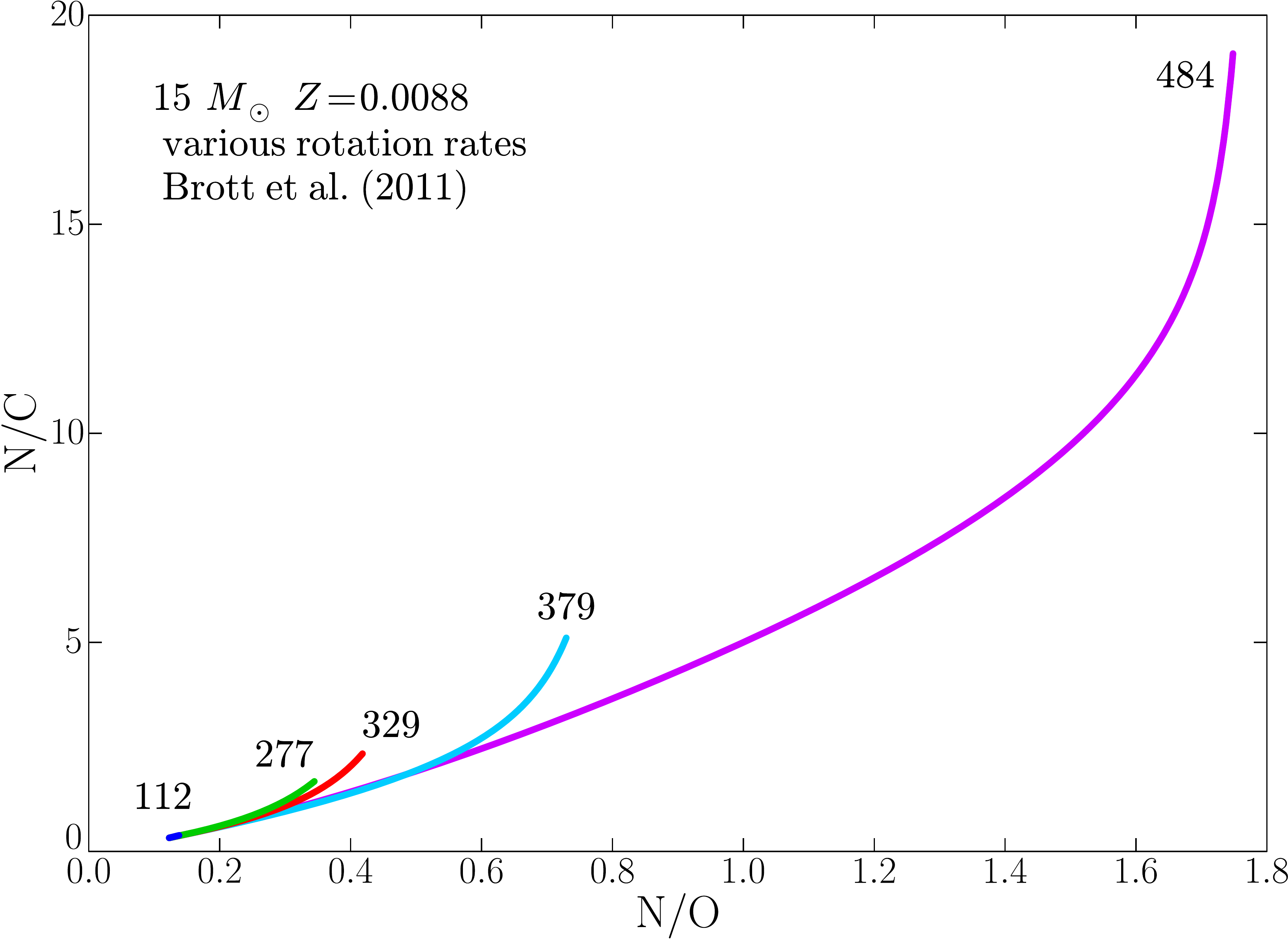}
\caption{The N/C vs~N/O abundances (in numbers) during the MS phase of a 
15\,$M_{\sun}$ model with $Z=0.0088$ \citep{Brott2011} with various initial
rotation velocities (in km\,s$^{-1}$). }
\label{BrottvMW}
\end{center}
\end{figure}

The effects of various initial rotational velocities in the N/C vs~N/O plot are illustrated for 15\,$M_{\sun}$~models in Fig.~\ref{M15014rot} \citep{GrilleBe} and in Fig. \ref{BrottvMW} \citepalias{Brott2011}. The models are labelled by their equatorial velocity\footnote{Note that in terms of $\Omega/\Omega_\text{crit}$ \citep[as labelled in][]{GrilleBe}, the 132 (277, 333, 396, and 501)\,km\,s$^{-1}$ model corresponds to 0.3 (0.6, 0.7, 0.8, and 0.9) respectively.}. The end of the MS phase is marked by a small open circle in Fig.~\ref{M15014rot}. The beginning of the red supergiant phase is indicated by a small square. The (first) blue supergiant phase extends between the open circle and the square. Figure~\ref{BrottvMW} only covers the MS of the models by \citetalias{Brott2011}. The behaviours of the two sets of models are different. For moderate velocities ($V_\text{eq}\lesssim400$\,km\,s$^{-1}$), as seen above, the N/O enrichments are higher in the Geneva models, while the more rapid rotators from \citetalias{Brott2011} are more enriched. For the N/C ratio, the trend is the same.

Another difference between the two sets of models is that at a given value of N/O, the Geneva models by \citet{GrilleBe} predict progressively higher N/C ratios for higher rotation velocities. The models by \citetalias{Brott2011} predict tracks in the N/C vs N/O plot with only small deviations from the average relation, with maybe a trend that is inverse of that in the Geneva models. We will see below that this trend, \textit{i.e.}, higher N/C ratios for \emph{lower} velocities at a given N/O value, is present in Fig. \ref{BrvelSMC} from the \citetalias{Brott2011} for the SMC composition. At this stage, the observations do not permit us to discriminate between the two sets of models.

 On the whole, we may conclude that, at least for low or moderate N/O values, the deviations from a linear relation remain small: in all cases, for the models by \citetalias{Brott2011}, we see that at N/O=0.4, 0.6, and 0.8, the maximum N/C values are limited to approximatively 2.2, 3.8, and 5.8, respectively, while for the Geneva models these values are 1.2, 2.4, and 4.4, respectively. The concept of a linear relation between N/C and N/O over a significant range also applies when the enrichments are produced by accretion of CNO-enriched material in a close binary (provided there is no differentiation brought by the dilution and transfer processes). This implies that discrepancies between the observations and the theoretical relations (at least in their initial parts) are difficult to explain with binary effects. This may be the reason why, even with binaries it is difficult \citepalias{Brott2011} to reproduce the CNO observations. Even when the magnetic braking is taken into account, which enhances the mixing \citep{BrakingMEM11}, the N/C vs N/O relation follows the same linear trend at the beginning.\\

%: fig 5
\begin{figure}[t]
\begin{center}
\includegraphics[width=.99\linewidth]{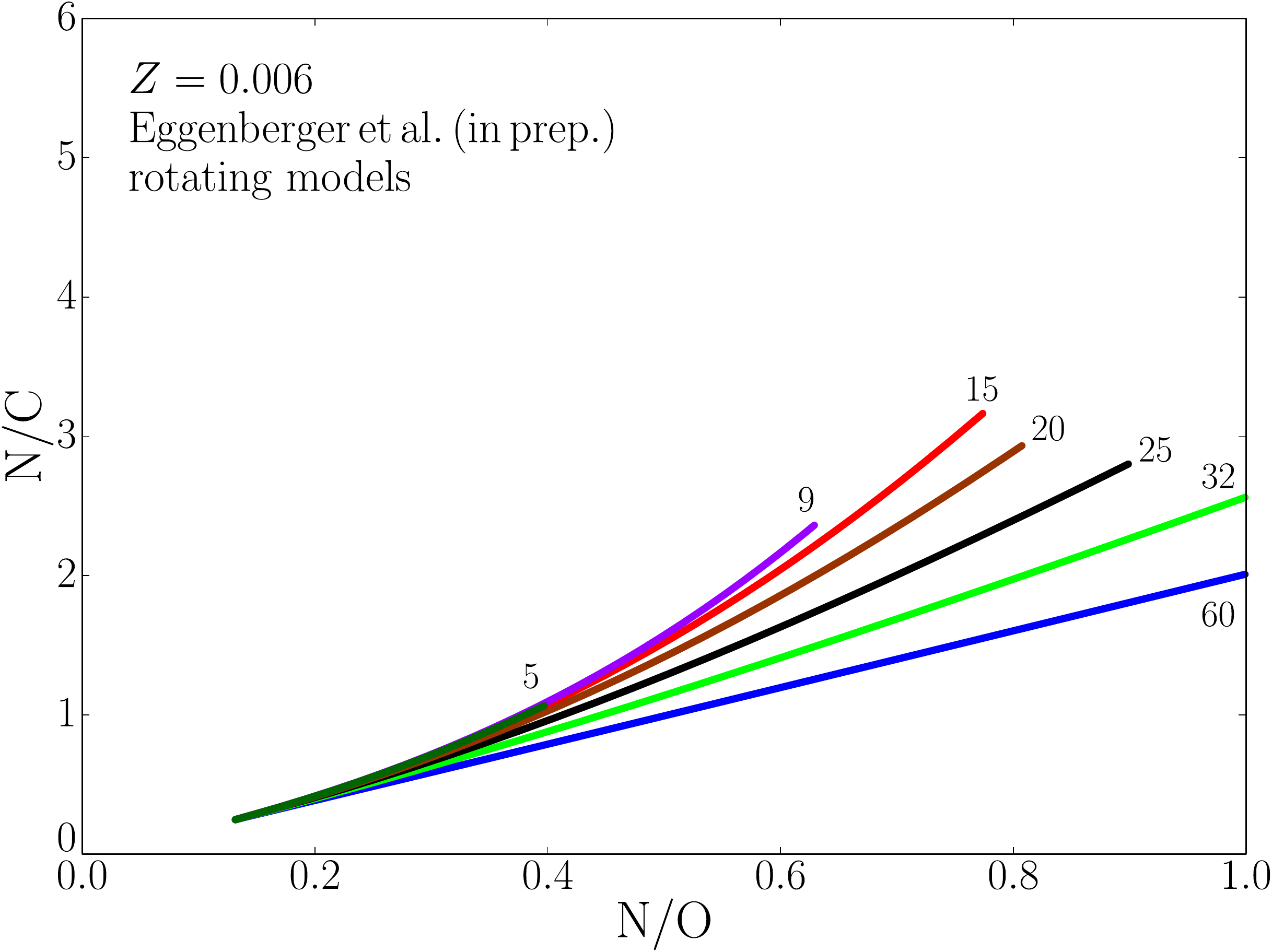}
\caption{The N/C vs~N/O abundances (in numbers) during the MS phase of models of rotating stars (Eggenberger et al., in prep.) for various stellar masses with $Z=0.006$. The models have initial velocities equal to 40\% of the critical value.}
\label{MODELMC}
\end{center}
\end{figure}

After the MS and the (first) blue supergiant phases, the models exhibit much larger effects of CNO processing. Large deviations from a linear relation may be present in the red supergiant phase ($\log(T_{\text{eff}})<3.7$) of models with high initial rotation velocities. For red supergiants, typical values of N/C are in the range of 2--15 and N/O values in the range of 0.7 -- 1.7. Red supergiants resulting from fast rotating MS stars have higher N/C enrichments with respect to their N/O values than initially slowly rotating stars. This is visible from Figs.~\ref{M15014rot} and \ref{M15002rot}. The extended convective zones of red supergiants homogenise the layers above the H-burning shell. This produces an average composition reflecting the already existing internal enrichments. These effects suggest that red supergiants, despite their generally very low rotation velocities, may show large differences in their N/C and N/O ratios depending on their rotation velocities during the MS phase.

When the stars enter the Wolf-Rayet stages, most of the composition differences disappear since these stars essentially show products of the CNO cycle at equilibrium, rather than a dilution of products of the CNO cycle in a medium of original composition. Note that some WNL stars may still show effects of dilution, while this is certainly not the case for stars in the WNE phase, which are nearly pure helium objects. In WN stars (WNL and WNE), the N/C ratios are typically in the range 10-100 and N/O between 5 and 80. Of course, in the WC stage, these two ratios rapidly fall to zero. 

%============================================================================
\subsection{Models for lower metallicities\label{Sec_lowZmod}}
%============================================================================

We use recent grids of numerical models of rotating stars for the metallicities of the Large and Small Magellanic Cloud (LMC and SMC) \citep[Eggenberger et al., in prep.]{Brott2011,GeorgyZ002}. Figures~\ref{MODELMC}, \ref{BrottLMC}, \ref{MODELSMC}, and \ref{BrottSMC} show the N/C~vs~N/O~plots for most of the MS phase of stars with initial masses in the range of 5 to 60 $M_{\sun}$ and for initial rotation velocities equal to $\sim$40 \% of the critical values. Again, without rotation there would be no N-enrichments during the MS phase for all masses considered here from 5 to 60\,$M_{\sun}$ included. From all models, we see, as already well known {\citep{MaederVII}, that at lower metallicities the mixing is stronger for a similar rotational velocity. This is due to steeper internal gradients of angular velocities $\Omega$ at lower $Z$, which favour a stronger diffusion. This results in more extended curves, as well as larger maximum deviations between the curves in the N/C~vs~N/O plots. 

%: fig 6
\begin{figure}[t]
\begin{center}
\includegraphics[width=\linewidth]{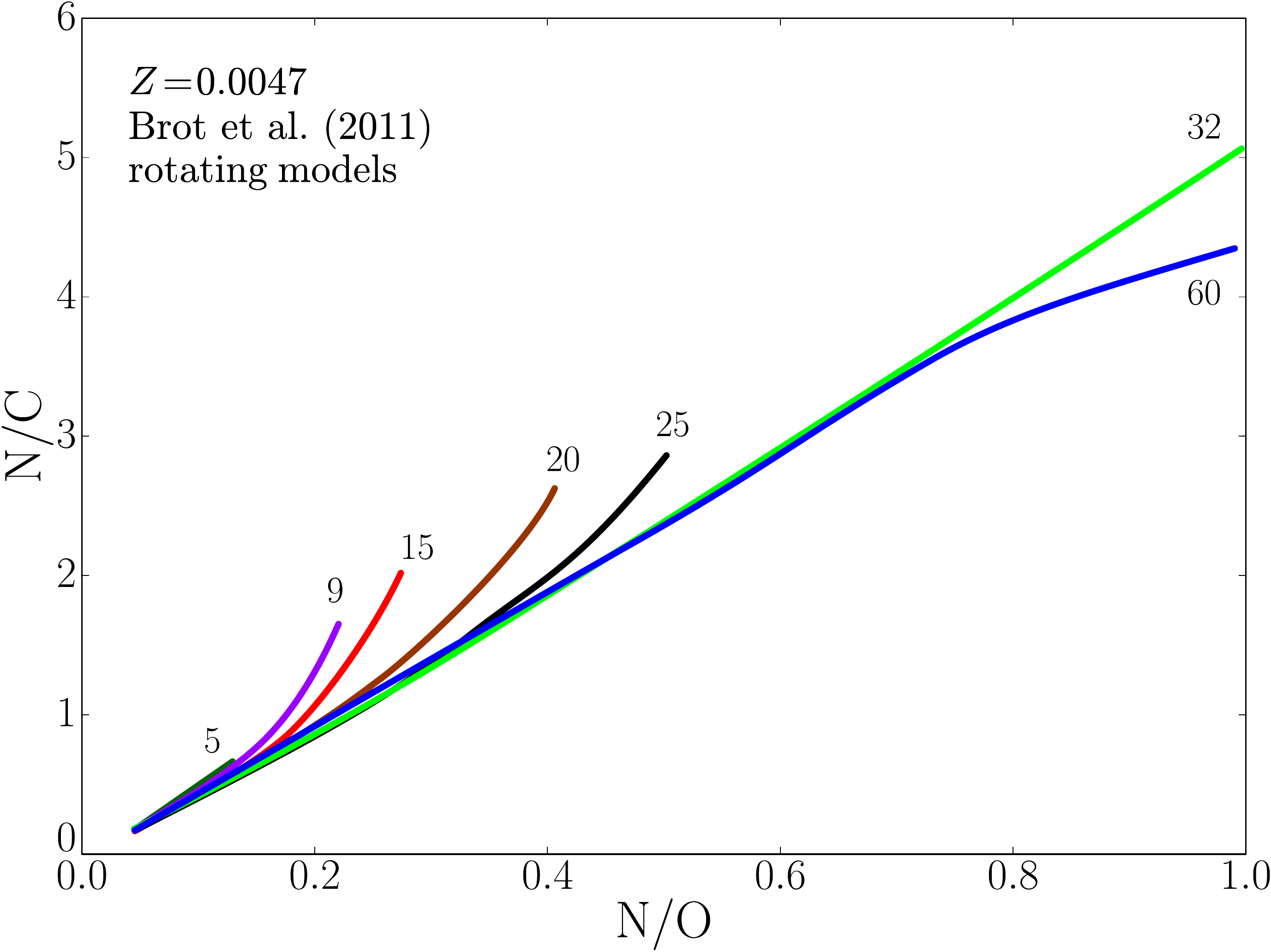}
\caption{The N/C vs~N/O abundances (in numbers) during the MS phase of models of rotating stars by \citet{Brott2011} for various stellar masses with metallicity $Z=0.0047$. Same remark as for Fig. \ref{MODELMC}.}
\label{BrottLMC}
\end{center}
\end{figure}

The models by \citetalias{Brott2011} use non-solar abundance ratios for the LMC and SMC, with nitrogen substantially under-abundant by $\sim$0.3 to 0.5\,dex (by number) relative to the other chemical species \citep[\textit{e.g.},][]{Garnett99,Kornetal02,Kornetal05,Rollestonetal03}. The initial N/C and N/O ratios are 0.141 and 0.035, respectively, for the LMC composition, and for the SMC they are 0.134 and 0.033. The Geneva models use solar CNO abundance ratios, which may apply to some low $Z$-galaxies with high star formation rates; their initial N/C ratio is 0.248 and the initial N/O ratio is 0.131. The differences with respect to the models by \citetalias{Brott2011} amount to a factor of about 1.8 for N/C and 3.8 for N/O for the LMC and SMC. These differences in initial CNO compositions have some consequences that require particular attention, because they make the comparison of models difficult:
\begin{enumerate}
\item There is a change in the zero points of the nuclear paths, visible as an offset between Figs. \ref{MODELMC} and \ref{BrottLMC}, and also between Figs. \ref{MODELSMC} and \ref{BrottSMC}.
\item The slopes of the relations N/C vs N/O are a bit steeper in the models by \citetalias{Brott2011} than in the Geneva models. This is partly due to the low initial N/O ratio, which decreases the denominator of the relations (\ref{highm}) and (\ref{lowm}).
\item The relative enhancements have to be estimated with accounts of the differences in the initial ratios. In contrast with what appears at first sight, the \emph{relative} mixing is a bit higher, at least for fast rotation, in the models by \citetalias{Brott2011} (Figs. \ref{BrottLMC}, \ref{BrottSMC}) than in the Geneva models (Figs. \ref{MODELMC}, \ref{MODELSMC}). This is a consequence of the different physics in the models (cf. Sect. 3.1).
\end{enumerate}

Both sets of models show that for low and moderate N/O ratios the differences between the curves corresponding to the average rotational velocities for various initial masses remain small. For example, in the Geneva models for $Z=0.006$ (Fig. \ref{MODELMC}) and $Z=0.002$ (Fig. \ref{MODELSMC}), the maximum N/C values are 1.15 and 2.2 at N/O= 0.4 and 0.6 (\textit{i.e.}, a relative N/O increase by more than a factor of 5); this is the same as for solar composition. In the models by \citetalias{Brott2011}, the deviations in Fig. \ref{BrottLMC} for the LMC are even smaller: up to N/O=0.5, \textit{i.e.}, a relative increase by a factor of 14, the N/C ratio remains smaller than 2.9. In Fig. \ref{BrottSMC}, the deviations are very limited up to N/O = 0.2, \textit{i.e.}, a relative increase of N/O by a factor of 6.

%: fig 7
\begin{figure}[t]
\begin{center}
\includegraphics[width=.99\linewidth]{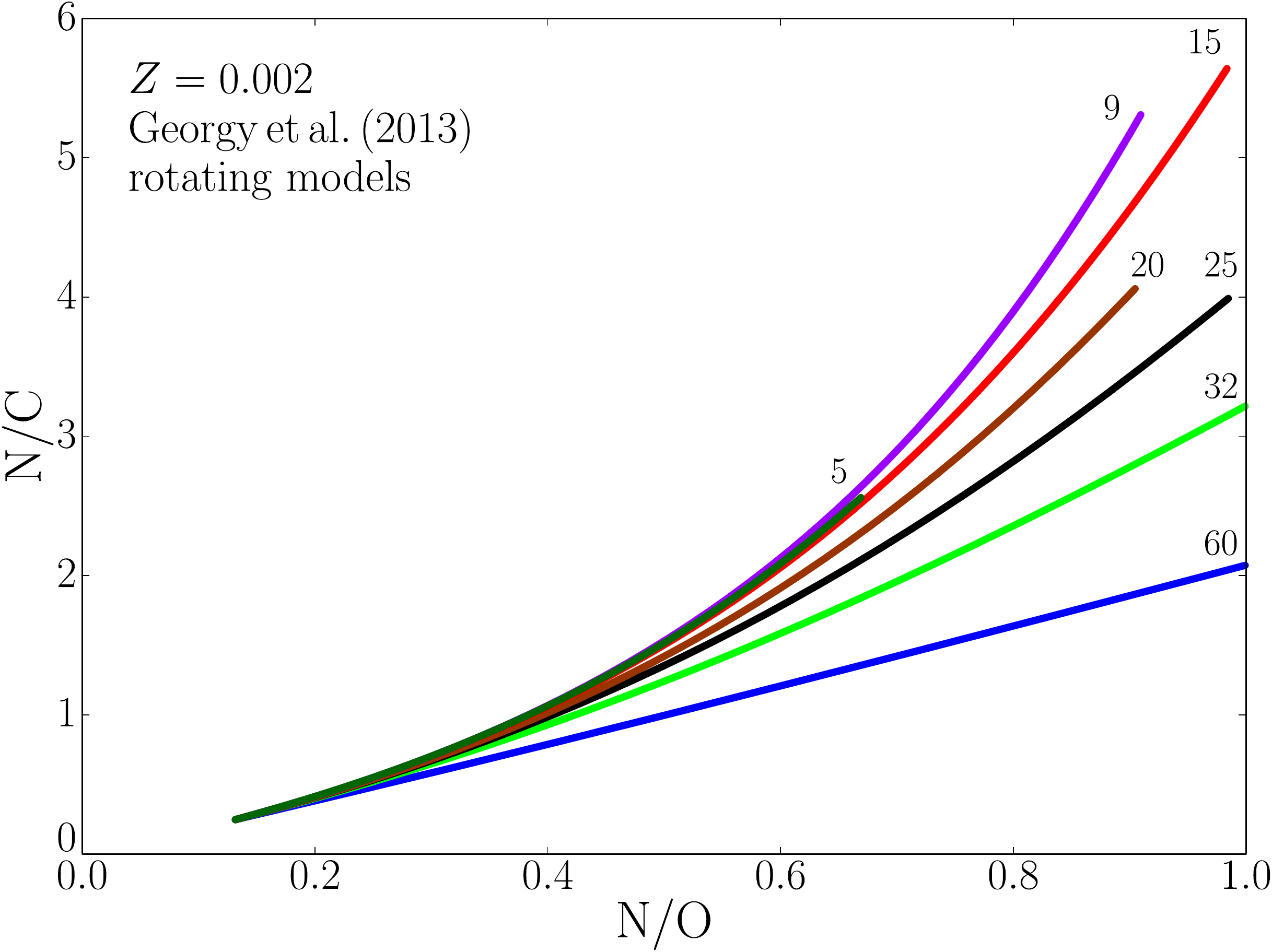}
\caption{The N/C vs~N/O abundances (in numbers) during the MS phase of models of rotating stars \citep{GeorgyZ002} for various stellar masses with metallicity $Z=0.002$. The models have initial velocities equal to 40\% of the critical value.}
\label{MODELSMC}
\end{center}
\end{figure}

%: fig 8
\begin{figure}[t]
\begin{center}
\includegraphics[width=.99\linewidth]{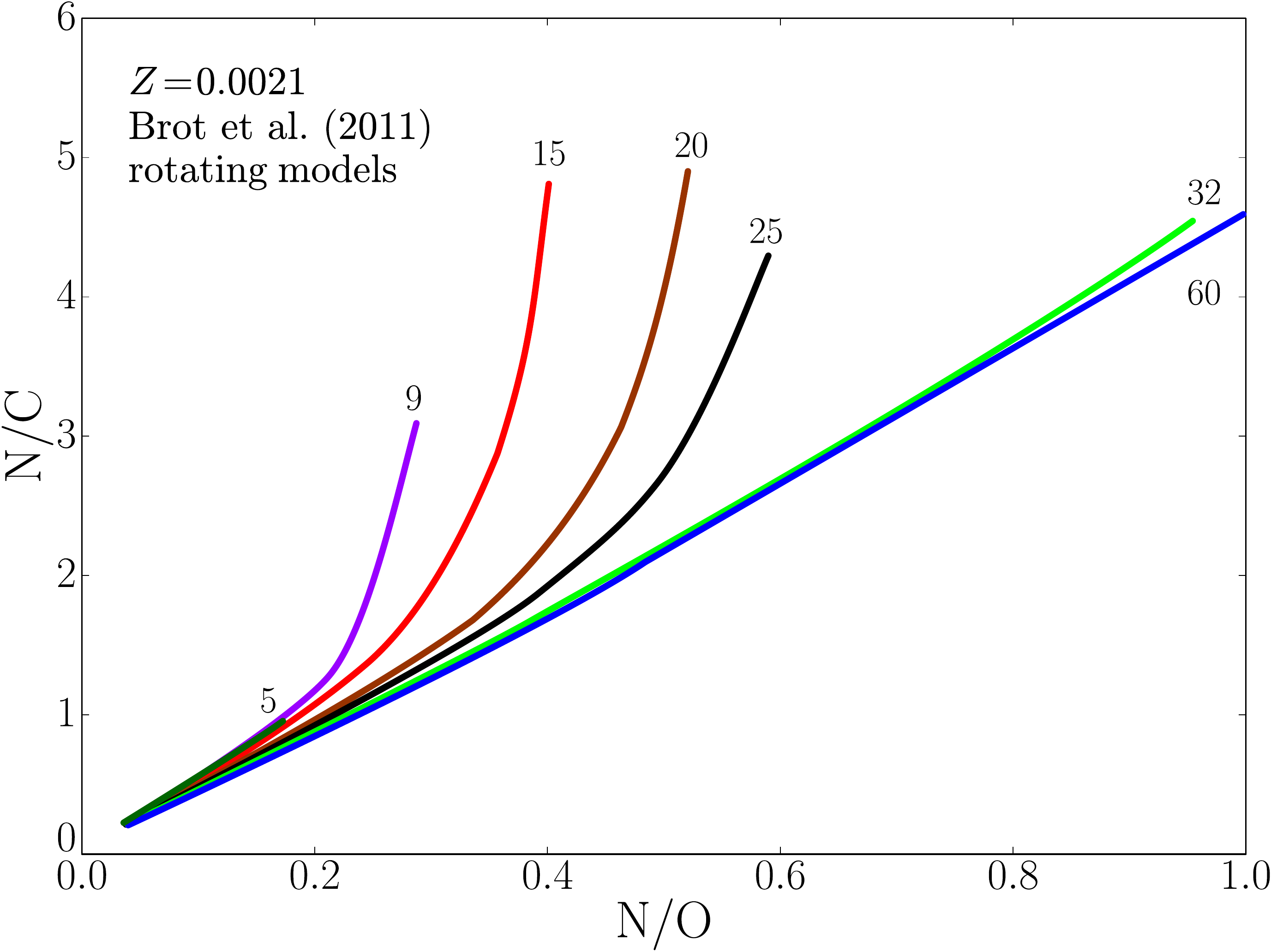}
\caption{The N/C vs~N/O abundances (in numbers) during the MS phase of models of rotating stars of various stellar masses with the compositions of the SMC \citep{Brott2011}. Same remark as for Fig. \ref{MODELMC}.}
\label{BrottSMC}
\end{center}
\end{figure}

On the whole, if one wants to use slightly different initial CNO ratios for already existing models of a given $Z$, the following rule applies to the first order: the relative enhancements with respect to the initial composition remain the same. As seen above, a slight change of the mean slope may also result in some cases.

%: fig 9
\begin{figure}[t]
\begin{center}
\includegraphics[width=.99\linewidth]{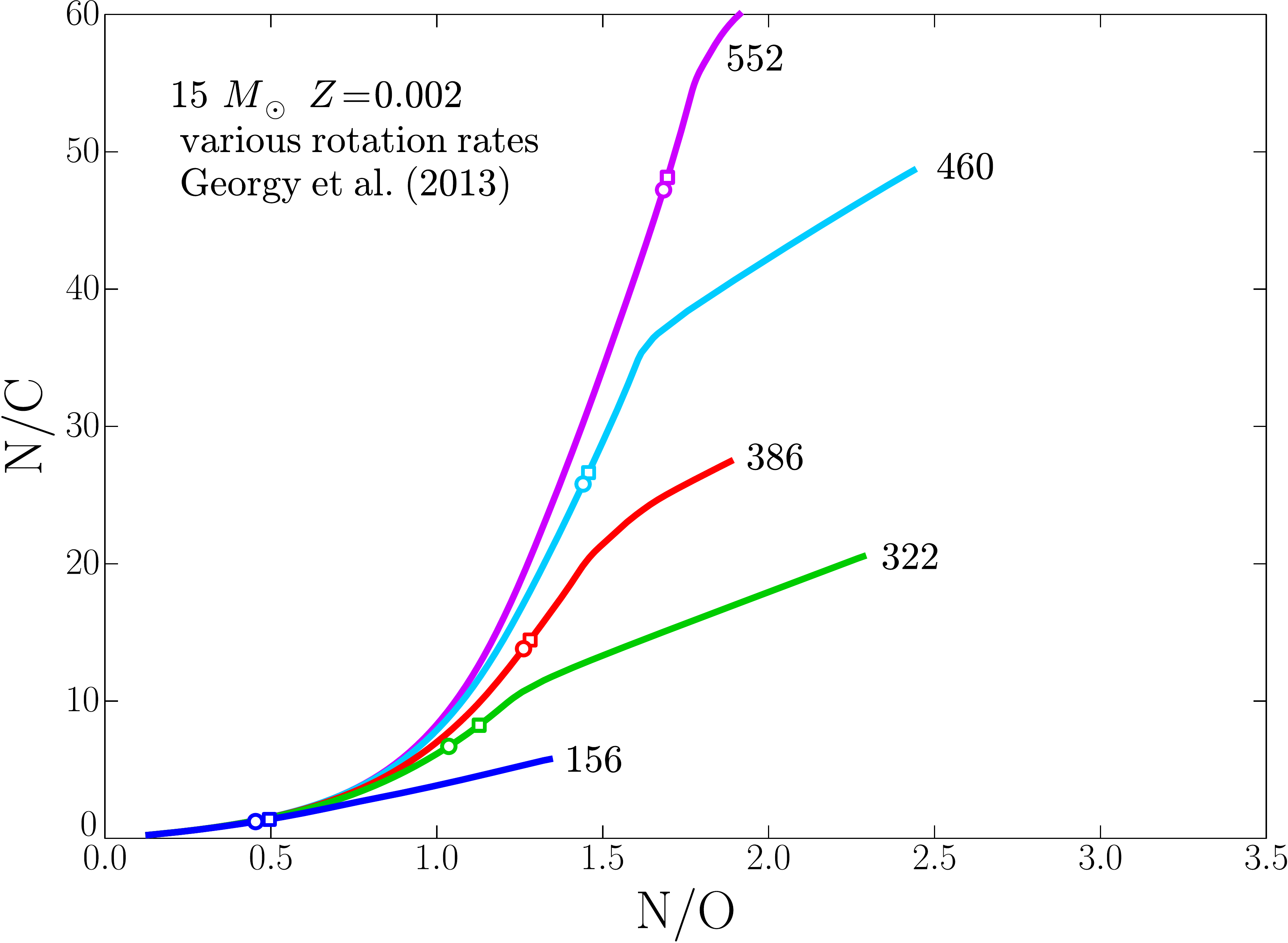}
\caption{Same as Fig.~\ref{M15014rot} but for $Z=0.002$ metallicity.}
\label{M15002rot}
\end{center}
\end{figure}

%: fig 10
\begin{figure}[t]
\begin{center}
\includegraphics[width=.99\linewidth]{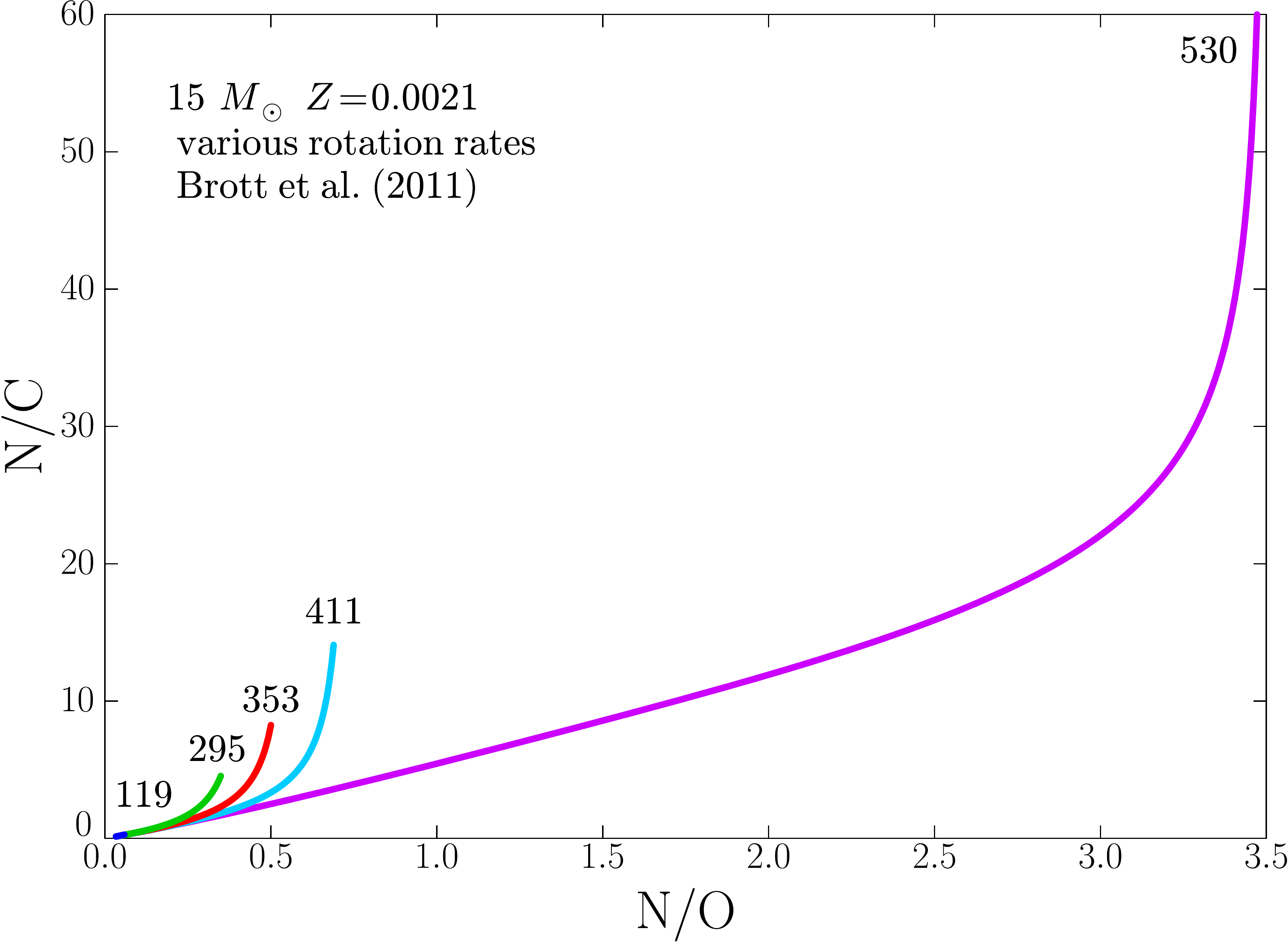}
\caption{Same as Fig.~\ref{BrottvMW} but for $Z=0.0021$ metallicity.}
\label{BrvelSMC}
\end{center}
\end{figure}

The effects of different initial rotational velocities for a given initial mass are also examined for the SMC metallicity. Figure~\ref{M15002rot} shows the results for the Geneva 15 $M_{\sun}$ models at $Z=0.002$ \citep{GrilleBe}. The N/C ratios can reach high values for high rotation velocities. However, up to N/O=0.8, the deviations from the unique linear relation remain small. At N/O=0.6 (relative enrichment larger than 5), the N/C ratio is below 2.4; at N/O=0.8 it lies below 4.4, even in the case of extreme rotation. This is similar to the solar metallicity case. Figure~\ref{BrvelSMC} shows the same kind of results for the models of a 15 $M_{\sun}$ star at $Z=0.0021$ for the SMC \citepalias{Brott2011}. We also notice high N/C and N/O values for high rotation velocities. The deviations from the linear relation remain moderate up to N/O= 0.30 (a relative increase by a factor of 9). There the maximum N/C ratio is of the order of 3. Physically, the N/C~vs~N/O curves express the degree of dilution of the CNO abundances at equilibrium mixed with the initial CNO abundances. All stellar models evolve along identical curves, at least in their initial linear part. Depending on model assumptions, models may evolve along these curves more or less rapidly.

Comparing the two sets of models and taking the difference of the zero points into account, we see that, at least for high velocities, mixing is more important in the models by \citetalias{Brott2011} than in the Geneva models for the different $Z$ values, which confirms the remark made in Sect. \ref{solarm}. There is another noticeable general difference between the two sets of models, in line with the previous remark on Figs. \ref{M15014rot} and \ref{BrottvMW}. The Geneva models at a given N/O ratio show higher N/C ratios for \emph{higher} initial rotational velocities, while at a given N/O value the models by \citetalias{Brott2011} show higher N/C values for \emph{lower} initial rotational velocities. This is a big difference, although not easy to test observationally. It likely results from a different history of mixing at various depths during MS evolution (cf. Sect. 2)}.

%%%%%%%%%%%%%%%%%%%%%%%%%%%%%%%%%%%%%%%%%%%%%%%%%%%
\section{Recent CNO observations in open clusters\label{comparisonobservation}}
%%%%%%%%%%%%%%%%%%%%%%%%%%%%%%%%%%%%%%%%%%%%%%%%%%%

There are several data sources for observations of CNO abundances in Galactic massive stars \citep{Przy2011}. Open clusters form, in principle, homogeneous samples in age and composition. Here, we focus on a work that is recognised as a seminal study of the massive star content of open clusters at different $Z$, in the Galaxy, the LMC, and SMC: the FLAMES Survey. Observations of over 800 stars were performed with the Fibre Large Array Multi-Element Spectrograph (FLAMES) on the 8.2\,m European Southern Observatory Very Large Telescope \citep[ESO-VLT,][]{Evansetal05,Evansetal06} and the main results of the chemical abundance analysis for about 300 stars were presented by \citet[hereafter H$+$09]{Hunteretal09}.\defcitealias{Hunteretal09}{H$+$09} The FLAMES Survey constitutes the largest sample of CNO abundances available in the three local galaxies and it has been used for various further analyses, particularly regarding the effects of stellar rotation on CNO abundances.

A warning is necessary here. The C-abundance is based on one \ion{C}{ii} line at 4267\,{\AA}, interpreted with an empirical adjustment. This is why for the comparisons made in Fig. 11-15, we indicate the error bars on the N/C ratios in the N/C vs~N/O plots of the various clusters studied in the Galaxy, LMC, and SMC. These plots were not used by \citetalias{Hunteretal09} to derive their results {\bf (instead, their conclusions were based on nitrogen abundances only)}. We also avoid stars for which only upper limits are given. However, despite these difficulties, these comparisons with the N/C vs N/O plots provide an independent quality test of the data and also offer some new indications on the possible deviations from model predictions and on the status of blue supergiants.

%: fig 11
\begin{figure}[t]
\begin{center}
\includegraphics[width=.99\linewidth]{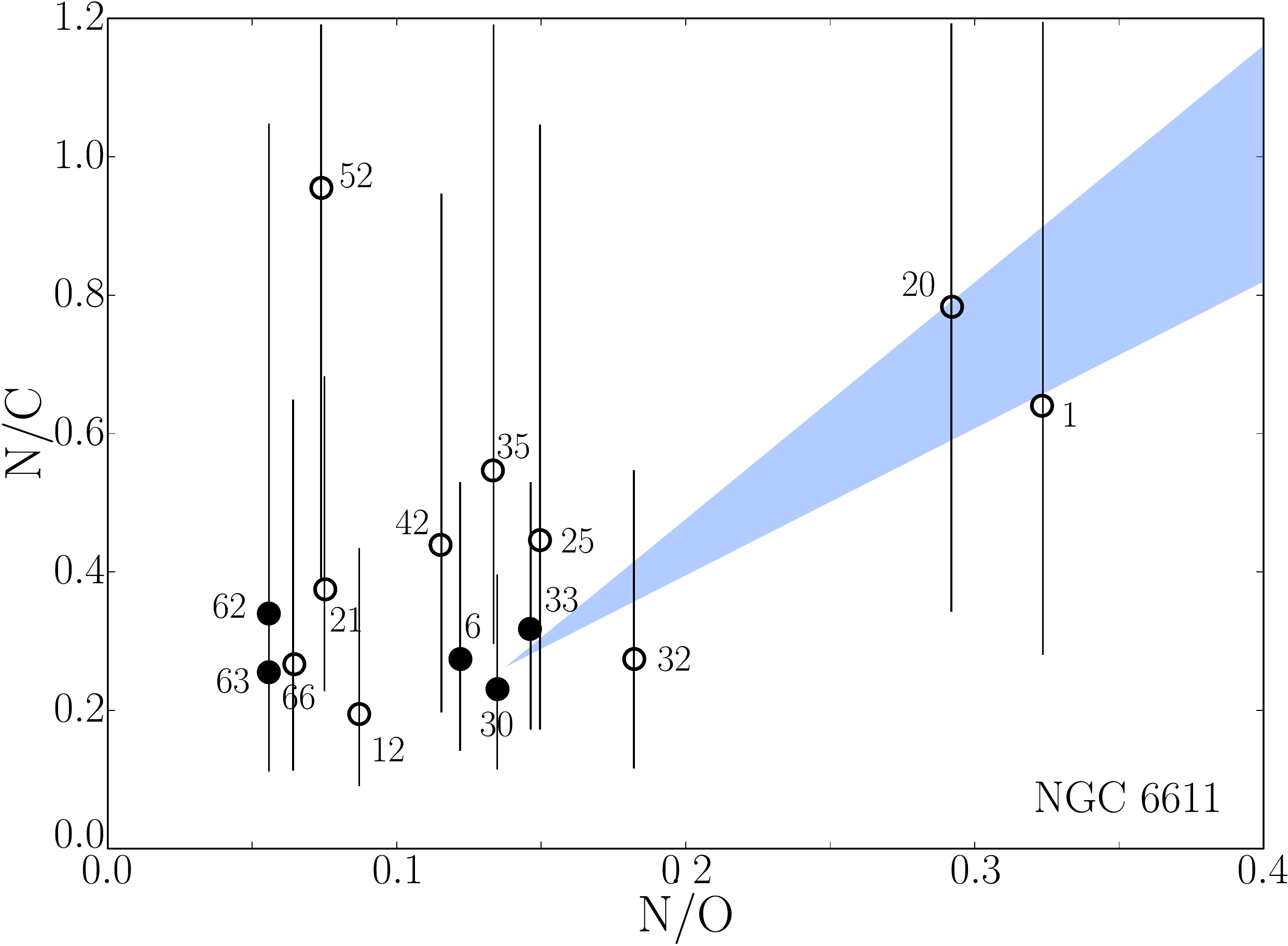}
\caption{The N/C vs~N/O abundances (in numbers) in the open cluster NGC\,6611 from observations by \citet{Hunteretal09}. The blue shaded area  shows the domain predicted for the Geneva stellar models, between 5 and 60 $M_{\sun}$ with solar composition (as given in Fig.~\ref{modelesNCNO}). The error bars on the observed N/C ratios are indicated. The error bars on the N/O ratios are of the same order of magnitude. The numbers attached to the data points correspond to the identifiers employed within the FLAMES Survey. Normal stars are marked by a filled symbol, binaries (or binaries candidates) or Be stars by an open symbol.}
\label{NGC6611}
\end{center}
\end{figure}

%============================================================================
\subsection{OB stars in the Galaxy}
%============================================================================

\begin{figure}[t]
\begin{center}
\includegraphics[width=.99\linewidth]{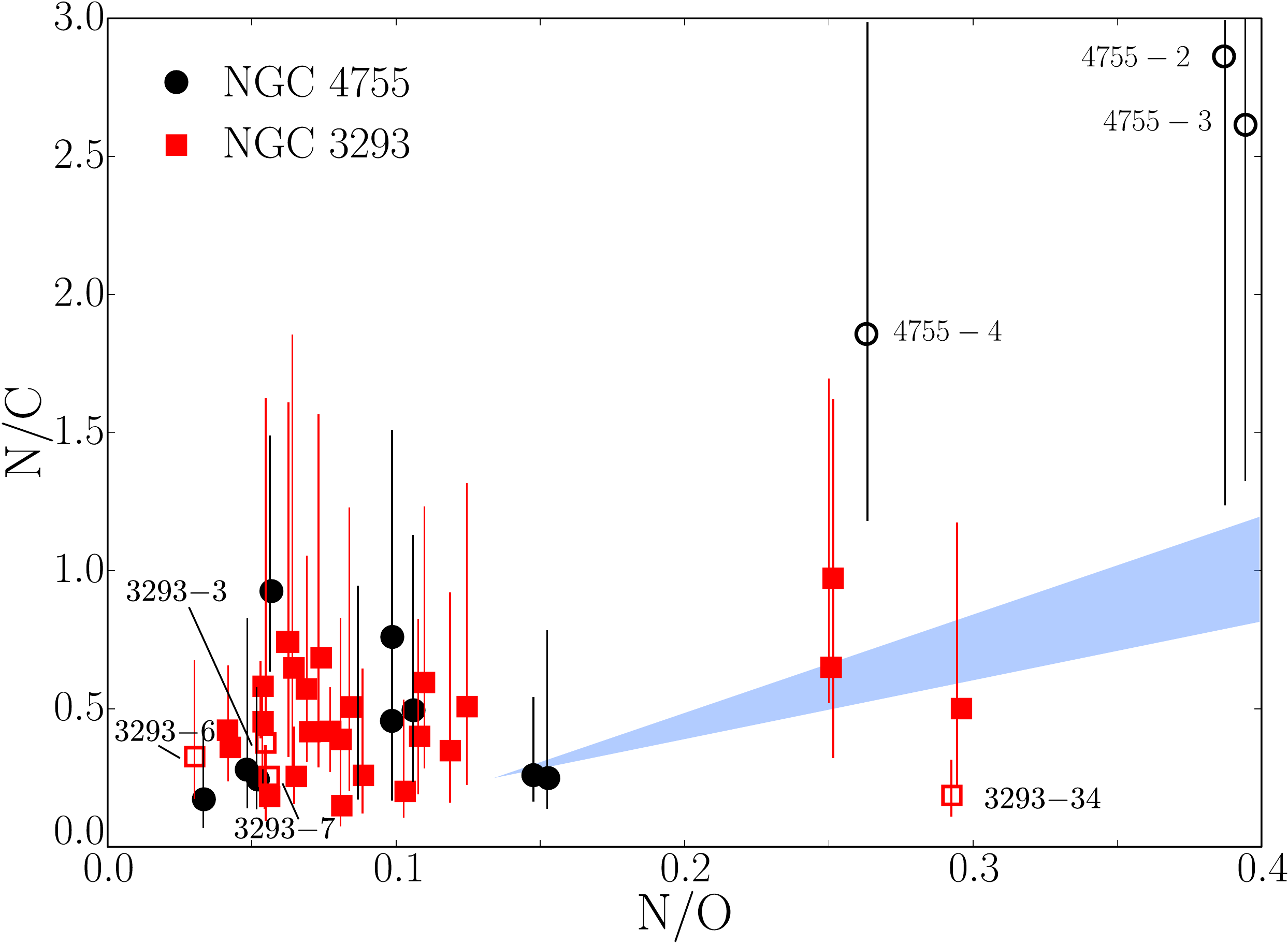}
\caption{The N/C vs~N/O abundances (in numbers) in the clusters NGC\,3293 and NGC\,4755 from data by \citet{Hunteretal09}. Same remark as for Fig.~\ref{NGC6611}. }
\label{NGC4755-3293}
\end{center}
\end{figure}

In the Galaxy, the three clusters NGC 6611 (7.7 Myr), NGC 3293 (10.3 Myr), and NGC 4755 (16.4 Myr) have been observed\footnote{The ages come from the Webda database, \url{http://www.univie.ac.at/webda}}. Figures~\ref{NGC6611} and \ref{NGC4755-3293} show the results of the comparisons between the domain covered by the Geneva models (Fig.~1) and observations. These figures show an offset of the objects with little or no enrichment with respect to the solar data given in Sect.~3 (N/C below 0.4 and N/O below 0.1). The offset can be understood as a consequence of the individual element abundances derived within the FLAMES Survey: while the carbon and nitrogen abundances are both about half solar, the oxygen abundance is only slightly sub-solar on average \citep[regarding solar values of][]{Asplundetal05,Asplundetal09}. Therefore, the offset in N/C is very small or even negligible, while the offset in N/O is much larger, with the observed values lower than the solar value. We refer for the further discussion of this to \citet{MaPa13} who have pointed out the unusual characteristics of the metallicity (about half solar) and, in particular, the C abundances found by the FLAMES Survey, when compared to several independent studies on Galactic stars.

The plot for NGC 6611 shows some scatter. However, most of the observed results are consistent with the model predictions from Fig.~1 indicated by a blue triangle in Fig.~\ref{NGC6611}. As discussed below, only stars 6, 30, 33, 62, and 63 (filled symbols in Fig.~\ref{NGC6611}) appear to be normal stars. The other stars are binaries or candidate binaries, and one star (52) is a Be star (all marked by open symbols in Fig.~\ref{NGC6611}). So, part of the observed scatter may be due to this fact.

The observed N/C enrichments for stars in NGC 4755 appear larger than for stars in NGC 3293, they are also significantly larger than the theoretical predictions. In principle, as stated previously, over the range considered here, the theoretical predictions should essentially be model independent. Thus, either the observed difference is a consequence of the scatter or it may indicate that the real N/C values are higher than predicted by both sets of models considered above and that rotational mixing makes higher N/C values than those we simulate in the models. The repetition below of such a trend is to be underlined. Let us also note that the re-analysis of seven stars (labeled in Fig.~\ref{NGC4755-3293}) shows that while stars NGC\,3293-3, 7, and NGC\,4755-2 do appear as normal stars, although with poor fits, the other four stars are binaries (NGC\,3293-6, NGC\,4755-3,4) or an He-strong star (NGC\,3293-34, see Table~\ref{indicators} and discussion in Sect.~\ref{assessment} and Appendix~B).

%: fig 12

%============================================================================
\subsection{OB stars in the LMC}
%============================================================================

\begin{figure}[t]
\begin{center}
\includegraphics[width=.99\linewidth]{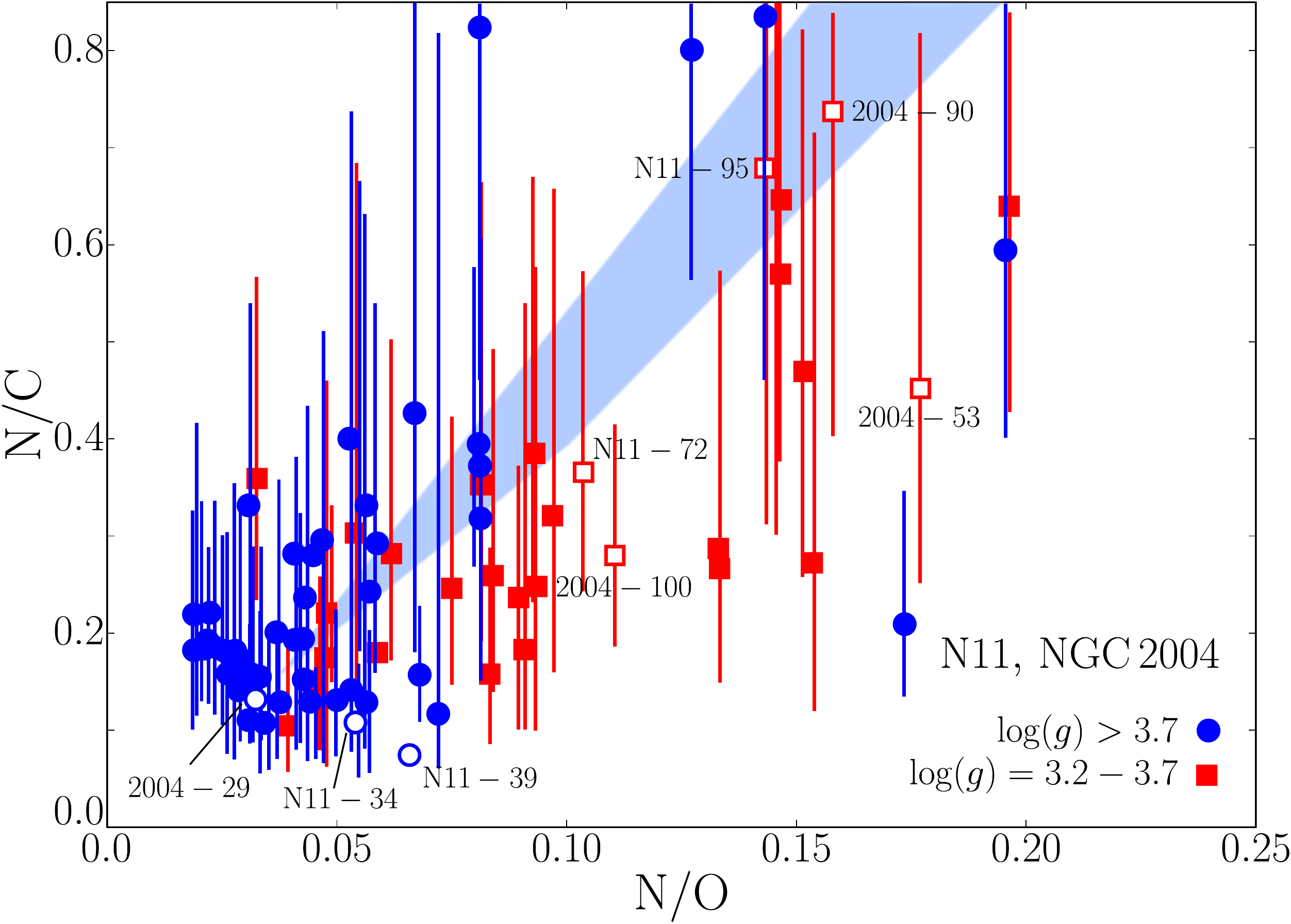}
\caption{The N/C vs~N/O abundances (in numbers) for the MS stars (stars with $\log g > 3.20$) in the LMC open clusters N\,11 and NGC\,2004 from observations by \citet{Hunteretal09}. The open symbols represent the stars given in Table~\ref{indicators}. The blue shaded area indicates the domain predicted by the models by \citet{Brott2011} of Fig. \ref{BrottLMC}. The initial N/C and N/O ratios are 0.141 and 0.035, respectively, as given in Sect.~\ref{Sec_lowZmod}. The error bars on the observed N/C ratios are indicated. The error bars on the N/O ratios are of the same order of magnitude.}
\label{N-11-2004-MS}
\end{center}
\end{figure}

We now turn to the two young clusters N11 and NGC 2004 in the LMC. The cluster N11 is about 8-10 Myr \citep{Evansetal06}, while NGC 2004 is older at of 25 Myr \citep{MaederGM99}. Figure~\ref{N-11-2004-MS} shows the comparison of the MS stars and the domain predicted by the models of Fig.~6 by \citet{Brott2011}. We see no significant offset between the abundances of the non-evolved models and those of the stars with negligible N-enrichments \citep{MaPa13}. The predicted trend is globally compatible with observations, however, the possibility of distinguishing between models like those discussed in Figs.~\ref{MODELMC} and \ref{BrottLMC} seems unlikely on the basis of the present data. The reanalysis of nine stars discussed below in Sect.~\ref{assessment} (see also Table~\ref{indicators}) indicate that only N11-95 and NGC\,2004-90 are normal stars, while the other seven stars are binaries (N11-8, 72) or Be stars (N11-34, 39, NGC\,2004-29, 53, and 100).

Figure~\ref{N-11-2004-supg} gives a broader view on the N/C vs N/O domain. The stars represented here are stars with $\log g \leq 3.20$. Noticeably, as we have seen above, some supergiant stars show higher N/C values compared to the theoretical predictions. Whether this is real or an effect of the scatter is uncertain. If real, this might either point towards many fast rotating stars or indicate an unaccounted effect in the mixing processes.

We point out that it is surprising to find several supergiant stars with very low values of N/O. Typically, see Fig.~\ref{BrottLMC}, one would expect that most blue supergiants would have N/O above 0.25-0.30, unless they would come from initial models with low initial velocities ($V_\text{ini}$ inferior to 40\% of the critical velocity). We may also wonder whether some stars are blue supergiants right after the MS phase, occupying the lower branch in Fig.~\ref{N-11-2004-supg} , while supergiants with high N/C values could be blue supergiants following the red supergiant stage. Thus, the FLAMES survey brings about several interesting questions, which may likely be answered in the future.
 
%: fig 13

%============================================================================
\subsection{OB stars in the SMC}
%============================================================================

In the SMC, the two clusters NGC 330 and NGC 346 are considered. Both are 20 Myr old, according to \citet{MaederGM99}. There seems to be two populations of stars in the latter cluster, one of very young stars of about 1\,Myr and another one of 20\,Myr \citep{deMarchi011}, which may form most of the present sample.

Figure~\ref{NGC330} confirms the results of previous figures. Also, we notice the absence of an offset between the observations and the models. Despite the scatter, there is much clearly evidences about internal mixing producing N enrichment at the expense of carbon and oxygen. We also note the existence of stars with high N/C values observed for rather low or moderate N/O ratios. Likely, such properties result from rapid rotation, however, it is still uncertain whether such excessively high N/C ratios indicate anomalous mixing or not. Nevertheless, the repetition of the same observed trend for all clusters considered in this work may be an indication in that direction.

%%%%%%%%%%%%%%%%%%%%%%%%%%%%%%%%%%%%%%%%%%%%%%%%%%%
\section{A re-assessment of the observational constraints\label{assessment}}
%%%%%%%%%%%%%%%%%%%%%%%%%%%%%%%%%%%%%%%%%%%%%%%%%%%

An underlying scatter much larger than one may expect from considering changes in the initial mass, rotation, and model details on the nuclear path is clearly evident in the N/O--N/C data. To assess the significance of the scatter, we aim to confirm the results of \citetalias{Hunteretal09} for a sub-sample of stars, both for objects following the nuclear path and for outliers. For the confirmation of the observational results we focus mostly on objects in the Galactic clusters (2/3 of the re-investigated sample) as spectra with the highest S/N were obtained for these. We further concentrate primarily on slow rotators ($V \sin i\,\lesssim\,100\,\text{km\,s}^{-1}$, 2/3 of the re-investigated sample), which facilitates uncovering details that are challenging to be identified in the lower-quality LMC and SMC data, and in faster rotators.

\begin{figure}[t]
\begin{center}
\includegraphics[width=.99\linewidth]{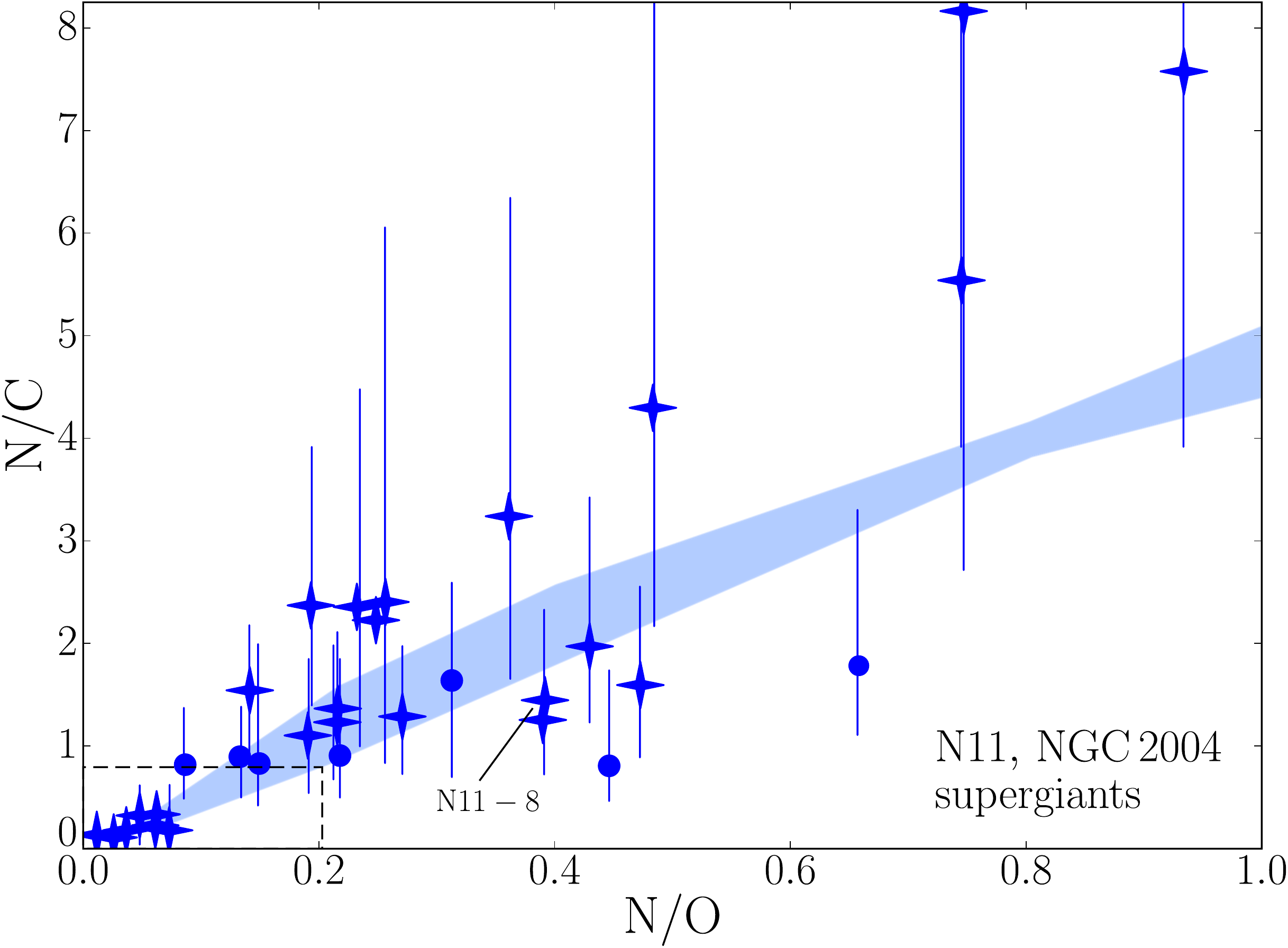}
\caption{The N/C vs~N/O abundances (in numbers) for stars with $\log g \leq 3.20 $, (blue supergiants: crosses; MS stars: circles) in the LMC open clusters N\,11 and NGC\,2004 from observations by \citetalias{Hunteretal09}. %\citet{Hunteretal09}.
Same remark as for Fig.~\ref{N-11-2004-MS}.}
\label{N-11-2004-supg}
\end{center}
\end{figure}

\begin{figure}[t]
\begin{center}
\includegraphics[width=.98\linewidth]{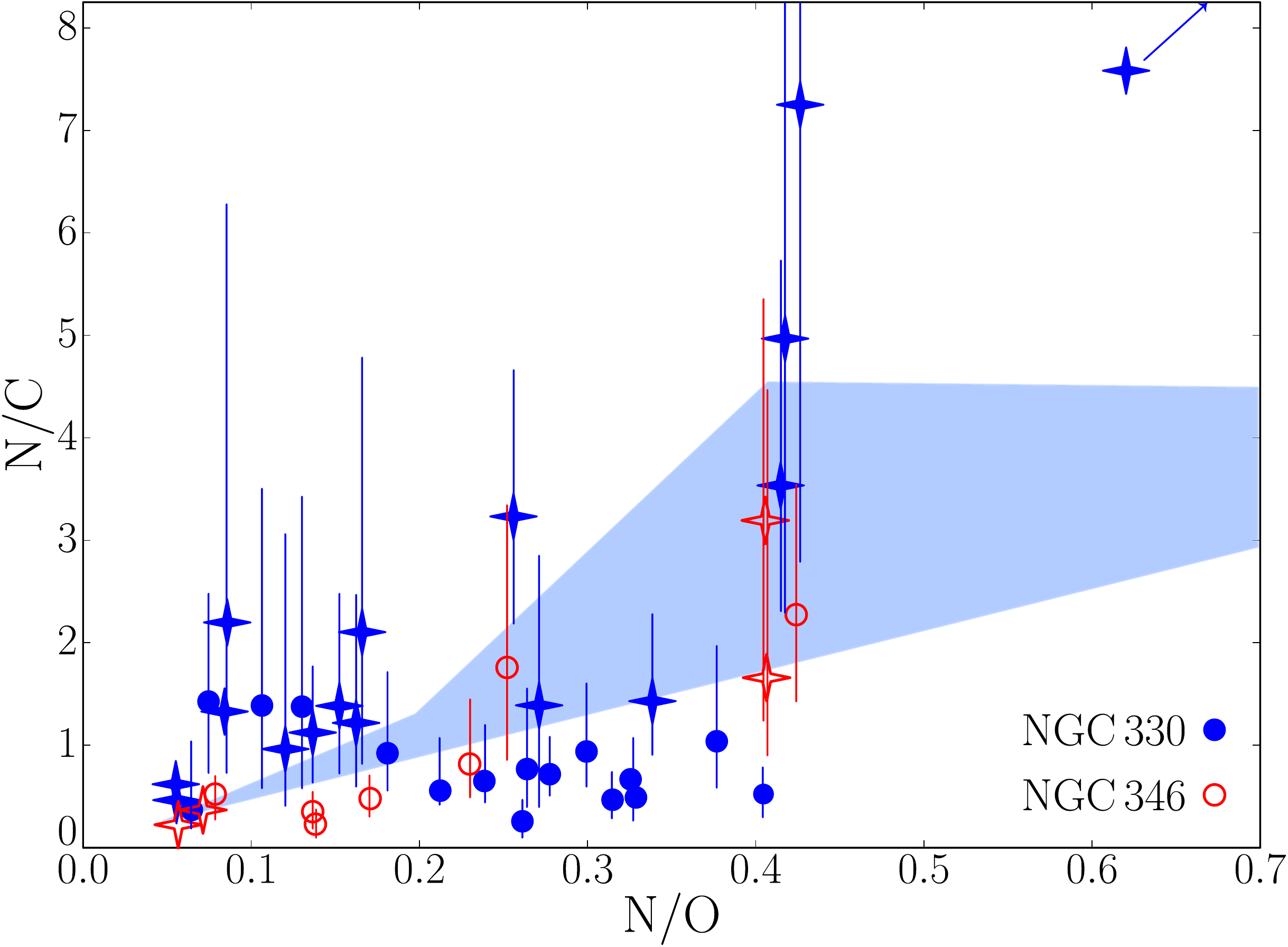}
\caption{The N/C vs~N/O abundances (in numbers) for stars in NGC\,330 (filled blue symbols) and NGC\,346 (open red symbols) in the SMC \citep{Hunteretal09}. Blue giants and blue supergiants are indicated by crosses and MS stars by circles. The star NGC\,330--004 with N/C=13.2 and N/O=0.955 is not indicated on this plot. The domain predicted by the models by \citet{Brott2011}~in Fig.~\ref{BrottSMC} is shown in blue. The initial N/C and N/O ratios are 0.134 and 0.033, respectively, as given in Sect.~2.2. The error bars on the observed N/C ratios are indicated. The error bars on the N/O ratios are of the same order of magnitude.}
\label{NGC330}
\end{center}
\end{figure}

In the following, we aim to investigate {\em all} of the results for one target cluster, NGC\,6611, to avoid any potential selection bias,. Then, additional targets were drawn from members of other clusters in the Milky Way and LMC, covering objects from the different star groups identified by \citetalias{Hunteretal09} (\textit{e.g.}, N-rich slow rotators and N-normal fast rotators on the MS, evolved stars). In total, 31 stars were selected, see Figs.~\ref{NGC6611} to \ref{N-11-2004-supg} for their identifications.

%: fig 14

A straightforward confirmation of the FLAMES Survey results of \citetalias{Hunteretal09} can be achieved under the premise that no matter the details of the model implementations, similar analysis results will be derived if they are sufficiently realistic. In particular, a good match of the synthetic with the observed spectra should be obtained globally, and in detail, as demonstrated by \citet[NP12]{NiPr12}. We therefore expect to find good agreement between the observed spectra and our model calculations when based on atmospheric parameters and abundances published by \citetalias{Hunteretal09}, except for small deviations that relate to details of the model implementations by \citetalias{Hunteretal09} and by our analysis (see Appendix~\ref{appendixA} for a short overview).

%: fig 15

Non-LTE synthetic spectra (\textit{i.e.}, accounting for deviations from local thermodynamic equilibrium) for the elements investigated by the FLAMES Survey were computed, employing atmospheric parameters and elemental abundances adopted from \citetalias{Hunteretal09}\footnote{Extracted from the online data table available via the VizieR Service at CDS: {\tt http://vizier.u-strasbg.fr/viz-bin/VizieR}}. Table~\ref{indicators} summarises the atmospheric input parameters effective temperature $T_\text{eff}$, surface gravity $\log g$, microturbulent velocity $\xi$, and (projected) rotational velocity $V\,\sin i$ for the selected sub-sample of stars. The computed synthetic spectra were convolved with a rotational profile for the appropriate $V\,\sin i$ and with a Gaussian profile of a width matching the spectral resolution to account for instrumental broadening. 

Comparisons with the observed spectra of the 31 sub-sample stars covered the diagnostic lines of \ion{H}{i}, \ion{C}{ii}, \ion{N}{ii}, \ion{O}{ii}, \ion{Mg}{ii}, and \ion{Si}{iii/iv} analysed by \citetalias{Hunteretal09}. In addition, we also verified the match of the \ion{He}{i/ii} and \ion{C}{iii} lines. The results of the visual inspection are summarised in Table~\ref{indicators}, assigning three quality indicators depending on whether a good, reasonable, or no match was found for the individual ions. Detailed examples are discussed in the online Appendix~\ref{appendixB}, with Figures~\ref{fit6611_006} to \ref{fit4755_003} showing the comparison between models and observations for six prototype cases. A good match is indicated when the deviations between model and observation are small overall, typical for what can be expected from a thorough investigation (see, \textit{e.g.},~most of the panels in Fig.~\ref{fit6611_006}, or the examples shown by NP12). For a reasonable fit, some lines of an ion may be over-reproduced and some others under-reproduced, but an average over the abundances from a full analysis would come close to the input abundance (see, \textit{e.g.},~the \ion{C}{ii} $\lambda$4267\,{\AA} and $\lambda\lambda$6578-82\,{\AA} lines in Fig.~\ref{fit3293_007}). No match is obtained when the modelled and observed spectral lines of an ion deviate significantly from each other, in particular, if the differences seem to be of a systematic nature (see, \textit{e.g.},~Fig.~\ref{fit4755_003}). The visual inspection of the observed spectra and the comparison with model spectra, furthermore, gave additional information on the nature of the re-investigated stars. Comments on this are summarised in the last column of Table~\ref{indicators}. Essentially four object classes were identified: normal stars, Be stars, chemically peculiar objects, and double-lined spectra. 

The following is learned from this comparison, with the relevant details discussed in Appendix~\ref{appendixB} on the basis of six prototype examples. It is rather easy to obtain an excellent fit to some individual spectral lines, and to obtain even a good fit to large portions of the observations. However, this becomes insignificant {\em if a single critical aspect is missed}, like an indicator for the presence of a disc, abundance peculiarities of an important chemical species, or the (often subtle) signatures for the presence of light contribution from a second star, as discussed in the following.\\[-8mm]

\paragraph{Binaries.} Line asymmetries and unresolvable difficulties in a consistent 
modelling of the line spectra point towards a certain or potential contribution of second 
light to 13 sample stars (see Table~\ref{indicators}). This is supported by findings, in 
particular for NGC\,6611, reported in the literature recently. Several sample stars have 
been identified as visual binaries with separations $\lesssim$1\arcsec using adaptive 
optics \citep{Ducheneetal01}, thus falling inside the FLAMES/MEDUSA fibre apertures of 
1.2{\arcsec}. The objects NGC\,6611-021, -025, and -066 show a magnitude difference between primary and secondary of $\sim$1{\fm}0, 5{\fm}3 and 3{\fm}7 in the $K$-band. This second light contributes to the NGC\,6611-021 spectrum (as confirmed by the observations), and possibly to the NGC\,6611-066 spectrum (a changing line asymmetry is indicated, but the S/N is low). Several other targets in NGC\,6611 were found to be radial velocity variables \citep{Martayanetal08}, from medium-resolution spectroscopy employing FLAMES/GIRAFFE on the VLT (confirmed by us). As the secondaries of OB-type primaries are often OB-stars as well, a contribution of second light to the spectrum is highly likely, as confirmed by us in several cases (see Table~\ref{indicators}). The number of multiple systems among the early B-star members, therefore, becomes comparable\footnote{The FLAMES Survey identified three out of 23 early B-stars as SB2 systems, \textit{i.e.},~a 13\%-fraction of binaries in NGC\,6611 \citep{Evansetal05}.} to the O-star binary fraction in NGC\,6611 \citep[44-67\%,][]{Sanaetal09}.
\begin{landscape}
\begin{table}[t!]
\centering
\caption[]{Verification of ionisation equilibria and abundance determinations and remarks on the sample stars.
\\[-6mm]\label{indicators}}
 \setlength{\tabcolsep}{.15cm}
 \tiny
 \begin{tabular}{rlrrrrccc@{\hspace{.7mm}}cc@{\hspace{.7mm}}ccccc@{\hspace{.7mm}}ccl}
 \noalign{}
\hline
\hline
  &               & \multicolumn{4}{c}{\citet{Hunteretal09}} & &   \multicolumn{12}{c}{our assessment of the \citetalias{Hunteretal09} work}\\
\cline{3-6} \cline{8-19} 
\# & Object        & $T_\text{eff}$ & $\log g$ & $\xi$      & $V \sin i$  & & H        &\ion{He}{i}&\ion{He}{ii}&\ion{C}{ii}&\ion{C}{iii}&\ion{N}{ii}&\ion{O}{ii}&\ion{Mg}{ii}&\ion{Si}{iii}&\ion{Si}{iv} & & remark\\
   &               & K                & (cgs)    & km\,s$^{-1}$ & km\,s$^{-1}$\\[-.1mm]
\hline\\[-2mm]
 1 & NGC\,6611-001 & 30300            & 3.70     & 10         & 142        & & $\bullet$ & $\circ$   & $\circ$   & $\circ$   & $\times$  & $\bullet$ & $\bullet$ & $\bullet$ & $\bullet$ & $\circ$   & & SB2\tablefootmark{a}, eclipsing binary\tablefootmark{b}\\%SB2
 2 & NGC\,6611-006 & 31250            & 4.00     &  8         &  20        & & $\bullet$ & $\bullet$ & $\circ$   & $\times$  & $\circ$   & $\circ$   & $\bullet$ & $\circ$   & $\bullet$ & $\bullet$ & & normal star\\
 3 & NGC\,6611-012 & 27200            & 3.90     &  5         &  95        & & $\bullet$ & $\bullet$ & $\times$  & $\times$  & $\times$  & $\circ$   & $\circ$   & $\circ$   & $\bullet$ & $\bullet$ & & SB2 candidate (asymmetric line profiles)\\%SB2
 4 & NGC\,6611-020 & 27000            & 4.10     &  2         & 220        & & $\bullet$ & $\bullet$ & $\times$  & $\times$  & $\times$  & $\circ$   & $\circ$   & $\bullet$ & $\bullet$ & --        & & SB\tablefootmark{c}, SB2 candidate (asymmetric line profiles)\\
 5 & NGC\,6611-021 & 26250            & 4.25     &  0         &  30        & & $\bullet$ & $\bullet$ & $\bullet$ & $\times$  & $\times$  & $\circ$   & $\times$  & $\bullet$ & $\bullet$ & $\circ$   & & visual binary\tablefootmark{d}, double-lined spectrum\\
 6 & NGC\,6611-025 & 25000            & 4.10     &  0         &  95        & & $\bullet$ & $\bullet$ & $\circ$   & $\times$  & $\times$  & $\bullet$ & $\circ$   & $\bullet$ & $\bullet$ & $\circ$   & & SB2 (changing asymmetry pattern), in addition: visual binary\tablefootmark{d}\\%SB2
 7 & NGC\,6611-030 & 22500            & 4.15     &  1         &  10        & & $\bullet$ & $\bullet$ & --        & $\times$  & $\circ$   & $\circ$   & $\circ$   & $\times$  & $\bullet$ & --        & & SB\tablefootmark{c}, SB1 (normal star)\\
 8 & NGC\,6611-032 & 22500            & 4.15     & --         &  70        & & $\bullet$ & $\bullet$ & --        & $\times$  & $\circ$   & $\bullet$ & $\bullet$ & --        & $\bullet$ & --        & & SB\tablefootmark{c}, SB2 candidate (asymmetric line profiles)\\
 9 & NGC\,6611-033 & 25600            & 4.00     &  4         &  30        & & $\bullet$ & $\bullet$ & $\bullet$ & $\times$  & $\bullet$ & $\circ$   & $\circ$   & $\circ$   & $\bullet$ & $\circ$   & & normal star\\
10 & NGC\,6611-035 & 27000            & 4.10     &  2         & 120        & & $\bullet$ & $\circ$   & $\times$  & $\times$  & $\times$  & $\bullet$ & $\bullet$ & --        & $\bullet$ & $\bullet$ & & double-lined spectrum (asymmetric line profiles)\\%SB2
11 & NGC\,6611-042 & 22500            & 3.90     &  4         & 155        & & $\bullet$ & $\circ$   & --        & $\bullet$ & $\circ$   & $\bullet$ & $\bullet$ & $\bullet$ & $\bullet$ & --        & & double-lined spectrum (asymmetric line profiles)\\
12 & NGC\,6611-052 & 19200            & 3.85     &  8         & 120        & & $\bullet$ & $\bullet$ & --        & $\circ$   & --        & $\bullet$ & $\bullet$ & $\times$  & $\bullet$ & --        & & Be star\\
13 & NGC\,6611-062 & 18400            & 4.05     &  7         & 165        & & $\bullet$ & $\bullet$ & --        & $\bullet$ & --        & $\bullet$ & --        & $\bullet$ & $\bullet$ & --        & & normal star\\
14 & NGC\,6611-063 & 22500            & 4.00     &  2         &  95        & & $\bullet$ & $\bullet$ & --        & $\bullet$ & --        & $\bullet$ & $\bullet$ & $\circ$   & $\bullet$ & --        & & SB\tablefootmark{c}, normal star? (status unclear because of low S/N)\\
15 & NGC\,6611-066 & 20800            & 4.20     & --         &  80        & & $\bullet$ & $\circ$   & --        & $\circ$   & --        & $\bullet$ & $\circ$   & $\times$  & $\bullet$ & --        & & SB2? (changing asymmetry pattern, but low S/N), also: visual binary\tablefootmark{d}\\
16 & NGC\,3293-003 & 20500            & 2.75     & 13         &  80        & & $\bullet$ & $\circ$   & $\times$  & $\times$  & $\circ$   & $\circ$   & $\times$  & $\times$  & $\circ$   & $\times$  & & normal star, overall poor fit, notable macroturbulence\\
17 & NGC\,3293-006 & 21500            & 3.15     & 15         & 200        & & $\bullet$ & $\bullet$ & $\bullet$ & $\bullet$ & --        & $\times$  & $\times$  & $\times$  & $\bullet$ & --        & & double-lined spectrum\\
18 & NGC\,3293-007 & 22600            & 3.10     & 11         &  65        & & $\bullet$ & $\times$  & $\circ$   & $\circ$   & $\times$  & $\times$  & $\times$  & $\times$  & $\times$  & $\times$  & & normal star, overall poor fit, notable macroturbulence\\
19 & NGC\,3293-034 & 26100            & 4.25     & --         &  35        & & $\bullet$ & $\times$  & $\bullet$ & $\times$  & $\times$  & $\bullet$ & $\bullet$ & $\times$  & $\circ$   & $\circ$   & & He-strong star\tablefootmark{e}\\
20 & NGC\,4755-002 & 15950            & 2.20     & 18         &  70        & & $\bullet$ & $\circ$   & --        & $\bullet$ & --        & $\circ$   & $\circ$   & $\times$  & $\bullet$ & --        & & normal star, notable macroturbulence, no ionization equilibrium\\
21 & NGC\,4755-003 & 17700            & 2.50     & 15         &  38        & & $\bullet$ & $\times$  & --        & $\times$  & $\bullet$ & $\times$  & $\times$  & $\times$  & $\times$  & $\times$  & & double-lined spectrum, overall poor fit\\
22 & NGC\,4755-004 & 19550            & 2.60     & 17         &  75        & & $\bullet$ & $\bullet$ & $\bullet$ & $\circ$   & $\times$  & $\circ$   & $\times$  & $\times$  & $\circ$   & $\times$  & & normal star, notable macroturbulence\\
23 & N11-008       & 25450            & 3.00     & 16         &  43        & & $\bullet$ & $\times$  & $\times$  & $\times$  & $\times$  & $\times$  & $\times$  & $\circ$   & $\times$  & $\times$  & & double-lined spectrum, notable macroturbulence, overall poor fit\\
24 & N11-034       & 25500            & 3.25     & 12         & 203        & & $\bullet$ & $\bullet$ & $\bullet$ & $\times$  & --        & $\circ$   & $\times$  & $\bullet$ & $\circ$   & --        & & Be star\\
25 & N11-039       & 21700            & 3.00     &  9         & 157        & & $\bullet$ & $\bullet$ & --        & $\times$  & --        & $\bullet$ & $\bullet$ & $\bullet$ & $\circ$   & --        & & Be star\\
26 & N11-072       & 28800            & 3.75     &  5         &  15        & & $\bullet$ & $\bullet$ & $\times$  & $\times$  & $\times$  & $\times$  & $\circ$   & $\circ$   & $\bullet$ & $\times$  & & double-lined spectrum\\
27 & N11-095       & 26800            & 3.85     &  2         & 267        & & $\bullet$ & $\bullet$ & --        & $\bullet$ & --        & $\bullet$ & $\bullet$ & $\times$  & $\bullet$ & --        & & normal star\\
28 & NGC\,2004-029 & 23100            & 3.50     &  1         &  30        & & $\circ$   & $\bullet$ & --        & $\circ$   & $\bullet$ & $\bullet$ & $\times$  & $\bullet$ & $\bullet$ & $\bullet$ & & Be star\tablefootmark{f}\\
29 & NGC\,2004-053 & 31500            & 4.15     &  7         &   7        & & $\bullet$ & $\bullet$ & $\bullet$ & $\times$  & $\circ$   & $\circ$   & $\circ$   & $\circ$   & $\bullet$ & $\circ$   & & Be star\tablefootmark{f}\\
30 & NGC\,2004-090 & 32500            & 4.10     & --         &  16        & & $\bullet$ & $\bullet$ & $\bullet$ & $\bullet$ & $\times$  & $\times$  & $\times$  & $\bullet$ & $\circ$   & $\circ$   & & normal star\\
31 & NGC\,2004-100 & 26800            & 3.70     &  6         & 323        & & $\bullet$ & $\bullet$ & --        & $\bullet$ & --        & $\bullet$ & $\times$  & --        & $\bullet$ & --        & & Be star\\
\hline
\vspace{-4mm}
\end{tabular}
\tablefoot{Symbol explanation:~~$\bullet$ good match, $\circ$ reasonable match, $\times$ no match, -- not observed. Remarks: this work, except
\tablefoottext{a}{\citet{Sanaetal09}}
\tablefoottext{b}{\citet{Lefevreetal09}}
\tablefoottext{c}{\citet{Martayanetal08}}
\tablefoottext{d}{\citet{Ducheneetal01}}
\tablefoottext{e}{\citet{{Evansetal05}}}
\tablefoottext{f}{\citet{{Evansetal06}}}
}
\end{table}
\end{landscape}
%
%: fig 16
\begin{figure}
\centering
\resizebox{0.99\hsize}{!}{\includegraphics[angle=0]{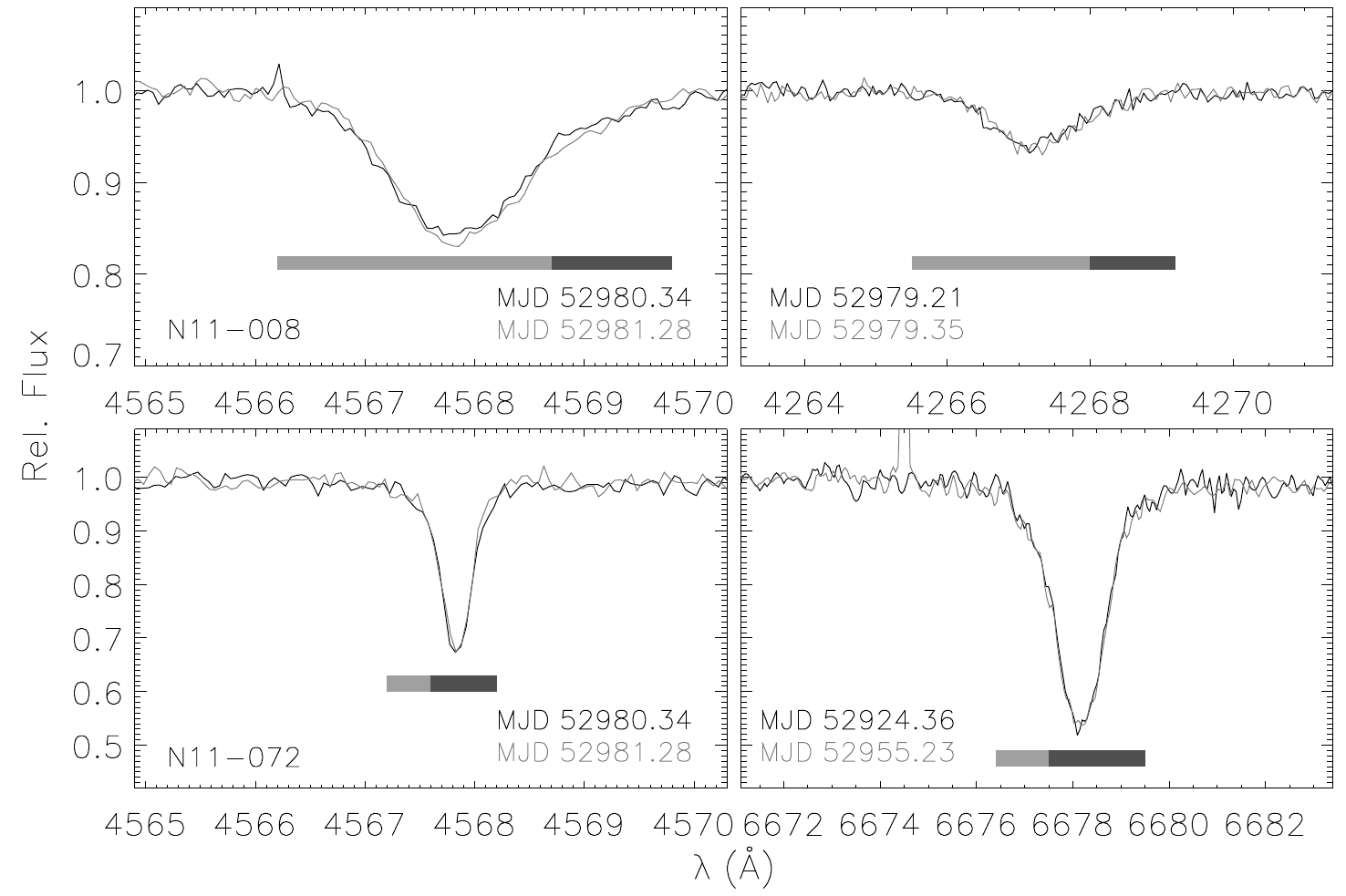}}
\caption[]{Examples of double-lined spectra within the LMC data of the FLAMES Survey. The spectral lines shown are \ion{Si}{iii}\,$\lambda$4567\,{\AA} and \ion{C}{ii}\,$\lambda$4267\,{\AA} for N11-008 (upper panels), and \ion{Si}{iii}\,$\lambda$4567\,{\AA} and \ion{He}{i}\,$\lambda$4713\,{\AA} for N11-072 (lower panels). Both objects are rv-stable within the timeline covered by the observations (epochs are indicated). The contributions of the components to the line profiles are indicated by grey bars. All spectra were cross-correlated with appropriate theoretical models and shifted to the laboratory rest-frame for illustration~purposes.}
\label{binaries}
\end{figure}

In general, the binary fraction of OB stars is high. A recent study by \citet{Chinietal12} 
finds a SB2-fraction among early B-type stars of $>$50\% in the mean, which may reach $>$70\% 
for stars in a young cluster. \citet{Sanaetal12} find similar numbers for the binary fraction 
of Galactic O-stars, 69$\pm$9\%, and a fraction of $>$50\% for the LMC Tarantula nebula 
region \citep{Sanaetal13}. Therefore, one may expect additional binary stars also in the 
LMC and SMC clusters of the FLAMES Survey to be identified, where so far a binary fraction 
of 23 to 36\% was found \citep{Evansetal06}. Examples of rv-stable double-lined spectra of 
LMC stars are shown in Fig.~\ref{binaries}. The lines are not clearly separated but 
tell-tale asymmetries reveal the true nature of the objects, which is immediately obvious for 
N11-008 and more subtle for N11-072. 

%
%: fig 17
\begin{figure}
\centering
\resizebox{0.99\hsize}{!}{\includegraphics[angle=0]{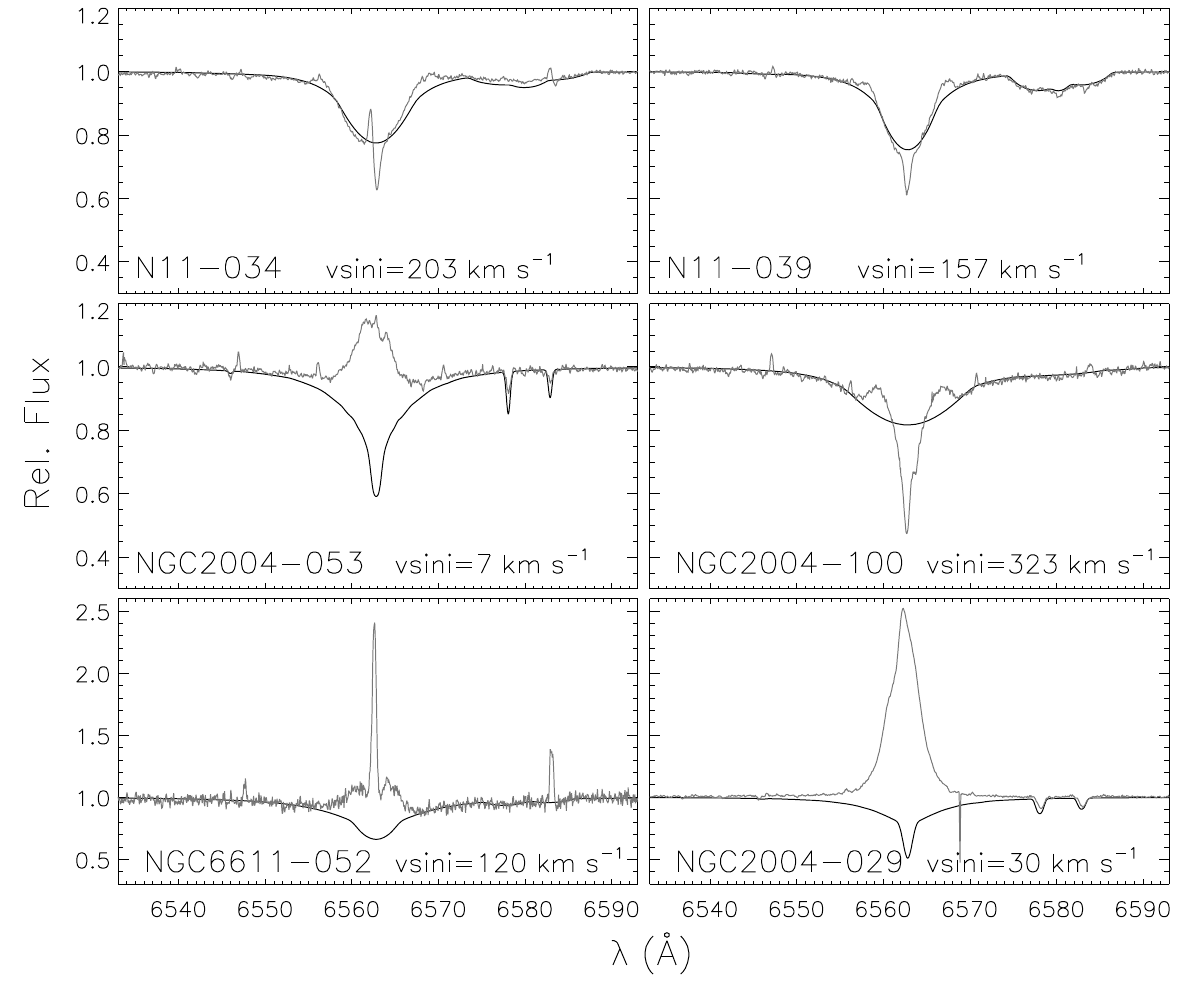}}
\caption[]{H$\alpha$ profiles of Be stars in our sub-sample of the FLAMES Survey. Grey lines show the observed spectra, black lines synthetic spectra based on parameters derived by the FLAMES Survey. Narrow features superimposed on the stellar lines are (residuals of corrected) nebular emission. Projected rotational velocities of the stars are indicated. Note the scale change of the ordinate in the lower two panels.}
\label{bestars}
\end{figure}

Overall, spectra influenced by second light need to be excluded from the analysis to avoid bias. The reason is that second light changes, depending on the individual binary configuration, line depths relative to the continuum flux and/or line widths, and thus also equivalent widths. An analysis of a binary composite spectrum, assumed it to be single, results in erroneous stellar parameters and elemental abundances.
\paragraph{Be stars.} A closer inspection of the spectra from our sub-sample of 31 stars selected from the FLAMES Survey also reveals Be characteristics for six stars, see Table~\ref{indicators}. When continuum and line emission from a disc overlaps with the stellar spectrum, this gives rise to a variety of spectral morphologies depending on disc parameters, disc size, and viewing angle on the star$+$disc system \citep[see, \textit{e.g.},][]{Riviniusetal13}. Moreover, Be-stars rotate close to the breakup velocity, leading to a distortion from spherical geometry and non-uniform surface temperatures (hot poles and cool equator) and densities. Gravitational darkening leads to an under-estimation of rotational velocities \citep{Townsendetal04} and to changes in the equivalent widths of the spectral lines \citep{Frematetal05}. Because of these complications to the modelling, all identified Be stars were excluded from analysis within the FLAMES Survey to avoid systematical bias to stellar parameters and elemental abundances. However, some known Be stars \citep[like NGC2004-029 and NGC2004-053,][]{Evansetal06} remained in the \citetalias{Hunteretal09} sample for unknown reasons, and we identified several more, including the fastest rotator of the FLAMES Survey, NGC\,2004-100. The H$\alpha$-profiles of these objects are shown in Fig.~\ref{bestars}. Note that the weak emission in H$\alpha$ line wings in the two objects in N11 may be inconspicuous to visual inspection, but clearly becomes apparent in the comparison with model predictions. It is not clear how many similar Be-stars may be present within the entire sample analysed by \citetalias{Hunteretal09} without performing detailed investigations.\\[-8mm]
\paragraph{Chemical peculiarity.} Furthermore, we identified one He-strong B-star in the sample analysed by \citetalias{Hunteretal09}: NGC\,3293-034 \citep[see also][]{Evansetal05}. The pronounced chemical peculiarity leads to a derivation of inappropriate stellar parameters and elemental abundances, if models with standard helium abundance are employed in the analysis.

Our re-assessment of a sub-sample of stars from the FLAMES Survey, as summarised in Table~\ref{indicators}, shows that problems from a standard analysis of the observational data may be widespread. A part of the analysis results of \citetalias{Hunteretal09} may therefore be subject to systematic bias of unknown extent. From the sub-sample of 31 stars re-investigated here only 11 seem suitable for a comparison of CNO values with theoretical predictions. When this is considered, {\em all} data points that are located off the nuclear path (as well as some matching ones) in Figs.~\ref{NGC6611} and \ref{NGC4755-3293} need to be dropped from interpretation (indicated by open symbols) and the scatter is reduced. 

Consequently, the main conclusion from our re-assessment is that a careful and critical re-investigation of the FLAMES Survey data presented by \citetalias{Hunteretal09} is clearly required. This may finally resolve the unusual characteristics of chemical abundances derived within the FLAMES Survey when compared to other studies of Galactic stars \citep{MaPa13}.

%%%%%%%%%%%%%%%%%%%%%%%%%%%%%%%%%%%%%%%%%%%%%%%%%%%
\section{Conclusions\label{conclusions}}
%%%%%%%%%%%%%%%%%%%%%%%%%%%%%%%%%%%%%%%%%%%%%%%%%%%

We provide some analytical properties of the N/C vs~N/O diagram. For small enrichments (less than a factor of 4 in N/O), these properties essentially depend on nuclear physics and initial CNO ratios. There are only small deviations from a straight line in the N/C vs~N/O diagram up to a value of N/O=0.6 at solar metallicity; for other compositions, this is the case up to an enrichment in N/O by a factor of at least 4. The N/C vs~N/O diagram, therefore, provides an interesting quality test for observational results.

The models of rotating stars by \citet{Brott2011} and the models by \citet{Grille2012} and \citet{GeorgyZ002,GrilleBe} show significant differences in their predictions. For low and moderate rotation velocities (up to 40\% of the critical value), the Geneva models predict larger enrichments, while for faster rotations the models by \citet{Brott2011} predict larger enrichments.

Observed CNO abundances have often been used to derive conclusions on the physical processes at work in massive stars, such as the occurrence of rotational mixing, mixing by magnetic instabilities, tidal mixing, instabilities, etc. Clearly, the data from the VLT-FLAMES Survey provide evidences of mixing. However, the precision of the data makes analyses of fine effects difficult. Also, the accuracy of the data is insufficient to allow for a test of the significant differences between the models of rotating stars by \citet{Brott2011} and the Geneva models. We note that the existence of stars with high N/C values for low or moderate N/O may be a bit difficult to explain in all models, but these findings may not be real.

A careful detailed re-assessment of a sub-sample comprising $\sim$10\% of the FLAMES Survey stars threw some light on sources of the above uncertainties. A double-lined character of many sample stars was found, as well as a chemical peculiarity or a Be-star nature for some stars. When the problematic cases are dropped from the interpretation, a much lower scatter of the observed data around the predicted nuclear path can be obtained.

We finally note that potentially, the N/C vs~N/O plot offers a powerful way to discriminate the blue supergiants before the red supergiant stage from those after it, but first the test of recovering the basic predictions of the nuclear CNO cycle must be passed satisfactorily.

\begin{acknowledgements}{Based on observations made with ESO Telescopes at the La Silla Paranal Observatory under programme IDs 171.D-0237 and 073.D-0234. We like to thank A.~Irrgang for help with the FEROS data reduction and M.A.~Urbaneja for providing us with test calculations employing {\sc Fastwind} and {\sc Cmfgen}. This research has made use of the SIMBAD database, operated at CDS, Strasbourg, France. MFN acknowledges financial support by the equal opportunities programme FFL of the University of Erlangen-Nuremberg.}
\end{acknowledgements}

%%%%%%%%%%%%%%%%%%%%%%%%%%%%%%%%%%%%%%%%%%%%%%%%%%%
\bibliographystyle{aa} % style aa.bst
\bibliography{aa20602}
%%%%%%%%%%%%%%%%%%%%%%%%%%%%%%%%%%%%%%%%%%%%%%%%%%%

\Online

\begin{appendix}

%%%%%%%%%%%%%%%%%%%%%%%%%%%%%%%%%%%%%%%%%%%%%%%%%%%
\section{Observational data and modelling}\label{appendixA}
%%%%%%%%%%%%%%%%%%%%%%%%%%%%%%%%%%%%%%%%%%%%%%%%%%%

We intend to work as closely as possible with the observational data of the FLAMES Survey, aiming to avoid bias from differences in the data reduction. As a test, raw data for all FLAMES Survey targets observed with the Fiber-fed Extended Range Optical Spectrograph (FEROS) at the 2.2m telescope at ESO (La Silla, Chile) were retrieved from the ESO archive and reduced using the FEROS pipeline and our additional recipes for continuum normalisation \citep[see, \textit{e.g.},][]{NiPr07}. These were compared with the published FLAMES Survey spectra, downloaded from the project webpage\footnote{\url{http://star.pst.qub.ac.uk}, see \citet{Evansetal05,Evansetal06} for details on the observations made within the FLAMES Survey and the data reduction.}. Good agreement of the two data reductions was found, such that the FLAMES Survey spectra were employed, except for the H$\alpha$ region, which was missing in the published data.

The sky-corrected single exposures of the individual FLAMES orders as well as the fully-reduced spectra were also downloaded from the project webpage. Multiple exposures of an object were cross-correlated to identify radial-velocity (rv) variables. In the case of rv variables, the single exposures per order were rv-corrected to laboratory rest frame, normalised and coadded (with weighing factors according to S/N), and then the orders merged. In the case of no rv-variation the fully-reduced spectra from the project webpage were employed.

Our atmospheric modelling and spectrum synthesis calculations for early B-type stars on the MS have been discussed in detail by \citet{NiPr08,NiPr12}. In brief, based on prescribed LTE model atmospheres \citep[{\sc Atlas9},][]{Kurucz93} non-LTE line-formation calculations were performed using the codes {\sc Detail} and {\sc Surface} \citep[both updated by K. Butler]{Giddings81,BuGi85}, abbreviated by {\sc Ads} henceforth. Such plane-parallel and hydrostatic hybrid non-LTE models have been shown to be equivalent to either plane-parallel and hydrostatic full non-LTE line-blanketed model atmospheres \citep{NiPr07,PNB11}, such as applied within the FLAMES Survey, or spherical and hydrodynamic full non-LTE line-blanketed model atmospheres \citep{NiSD11} on the MS. 

%%%%%%%%%%%%%%%%%%%%%%%%%%%%%%%%%%%%%%%%%%%%%%%%%%%
\section{Test case examples}\label{appendixB}
%%%%%%%%%%%%%%%%%%%%%%%%%%%%%%%%%%%%%%%%%%%%%%%%%%%

\begin{figure*}
\centering
\resizebox{0.89\hsize}{!}{\includegraphics[angle=0]{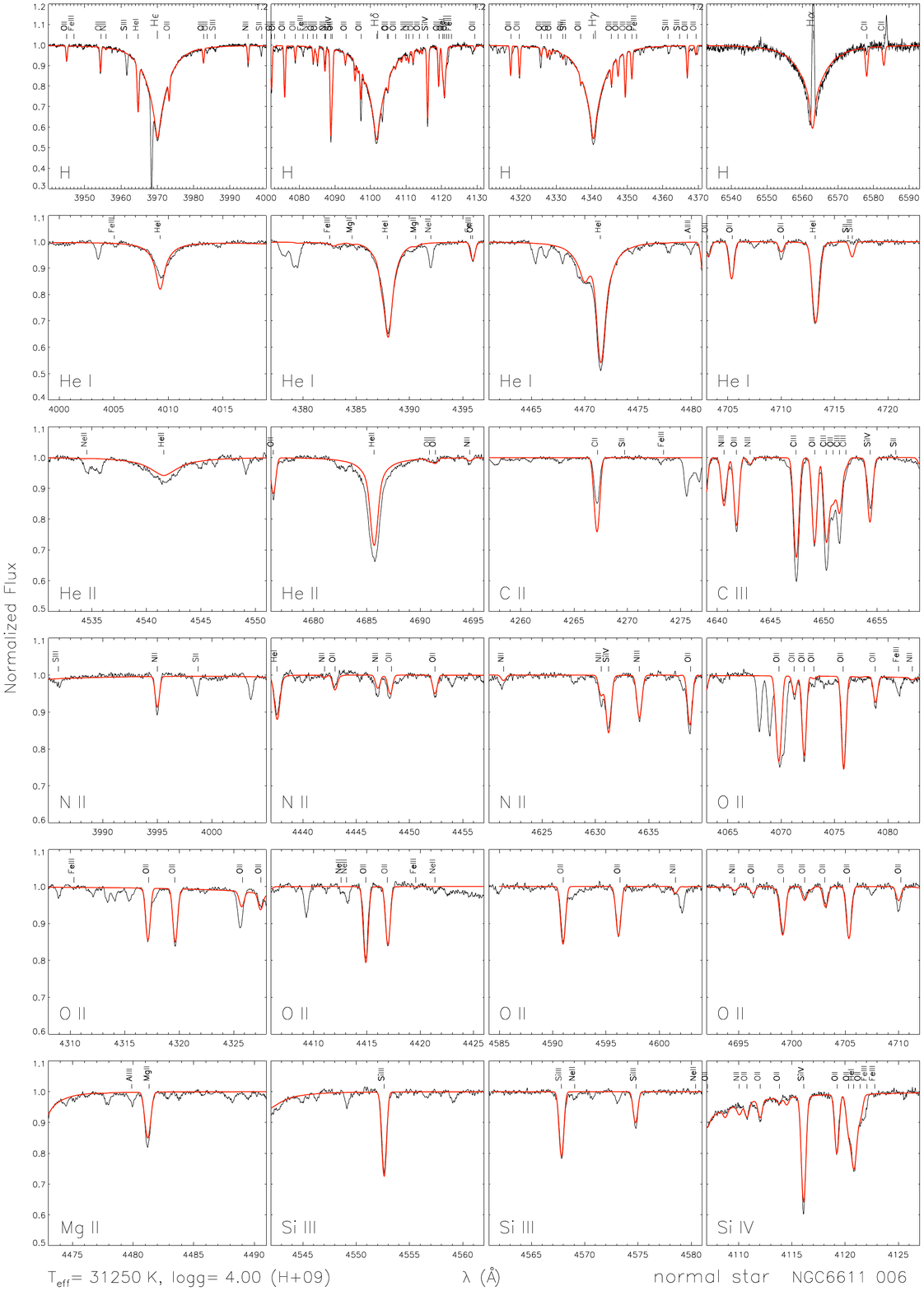}}
\vspace{-.5cm}
\caption[]{Comparison of a synthetic spectrum computed by us with {\sc Ads} for the atmospheric parameters and elemental abundances derived by \citet{Hunteretal09} (red line) and the {\sc Flames/Giraffe} observation (black line) for the star NGC\,6611-006. Displayed are spectral regions with important features for the analysis, as indicated. The sharp H$\alpha$ emission feature is of nebular origin. This is an example of one of the best matches between theory and observation found within our re-investigation of the sub-sample of stars drawn from the \citet{Hunteretal09} work.}
\label{fit6611_006}
\end{figure*}

\begin{figure*}
\centering
\resizebox{0.89\hsize}{!}{\includegraphics[angle=0]{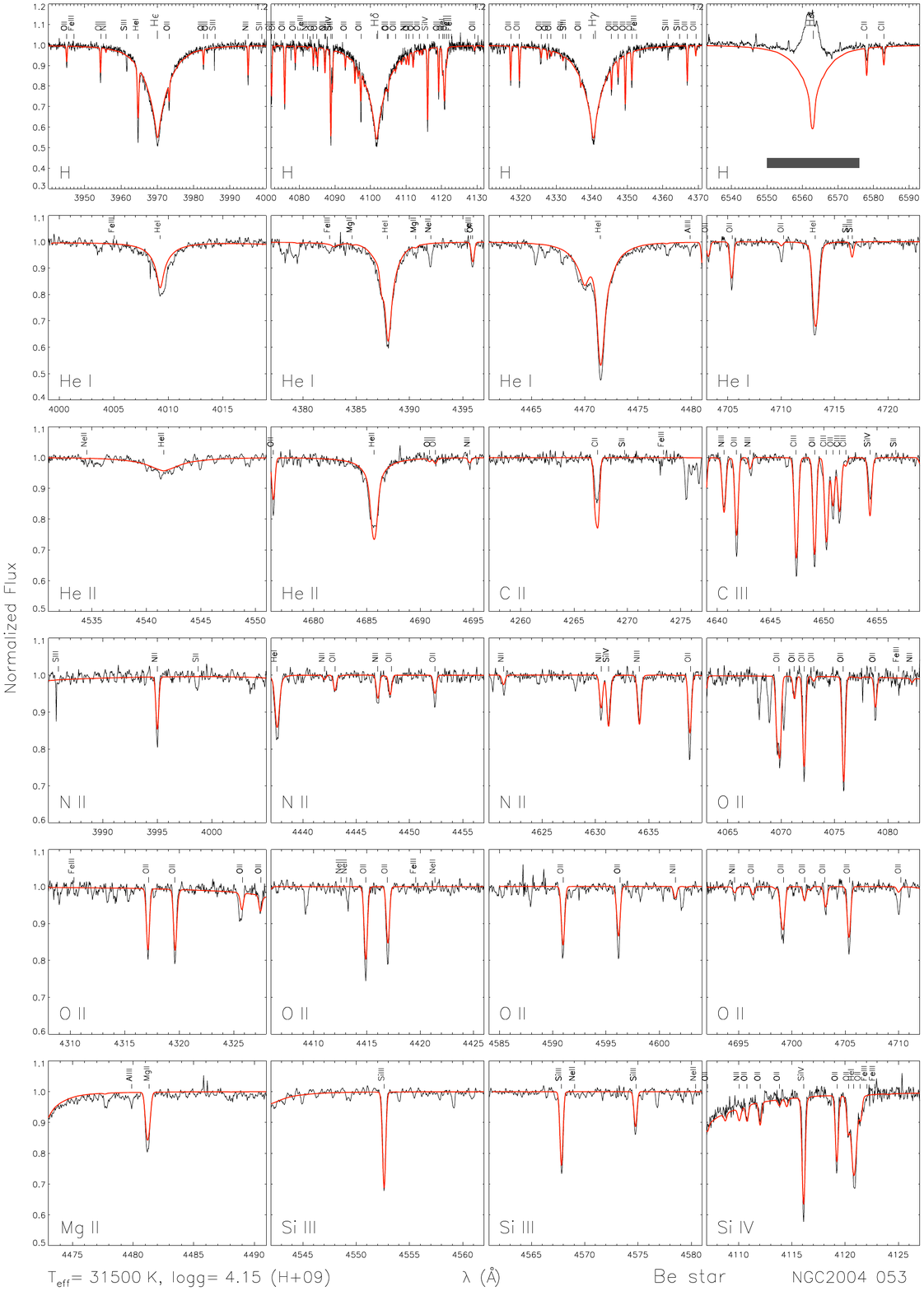}}
\vspace{-.5cm}
\caption[]{Same as Fig.~\ref{fit6611_006}, but for the star NGC\,2004-053. This is an example for a model fit to a Be star seen pole-on. The grey-shaded box guides the eye to recognise the signature of the presence of a disc, \textit{i.e.}, the H$\alpha$ emission. Similar to the models employed by \citet{Hunteretal09}, the present computations with {\sc Ads} do not account for the presence of a disc and gravity darkening. The neglect of the true nature of this star yields potentially biased atmospheric parameters and abundances, despite the fact that a good fit to many spectral lines is achieved.}
\label{fit2004_053}
\end{figure*}

\begin{figure*}
\centering
\resizebox{0.89\hsize}{!}{\includegraphics[angle=0]{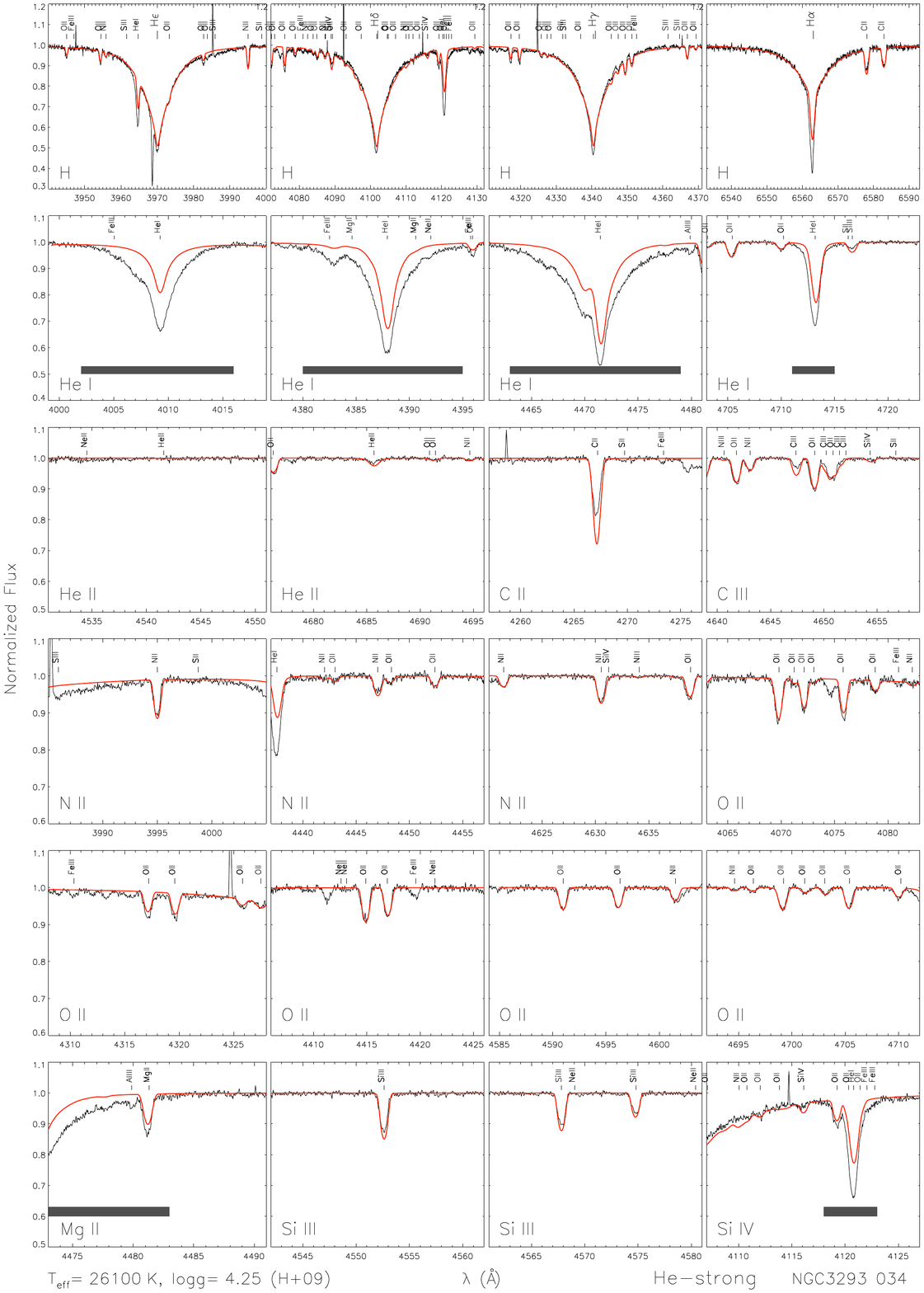}}
\vspace{-.5cm}
\caption[]{Same as Fig.~\ref{fit6611_006}, but for the He-strong star NGC\,3293-034. The grey-shaded boxes guide the eye to recognise the signatures of the chemical peculiarity of the star, \textit{i.e.}, the unusually strong \ion{He}{i} lines. Similar to the models employed by \citet{Hunteretal09}, the present computations with {\sc Ads} do not account for the pronounced helium enrichment and therefore the altered atmospheric structure of the star. As a consequence, the analysis yields incorrect atmospheric parameters and abundances, despite the fact that a good fit to the hydrogen and many metal lines is achieved.}
\label{fit3293_034}
\end{figure*}

\begin{figure*}
\centering
\resizebox{0.89\hsize}{!}{\includegraphics[angle=0]{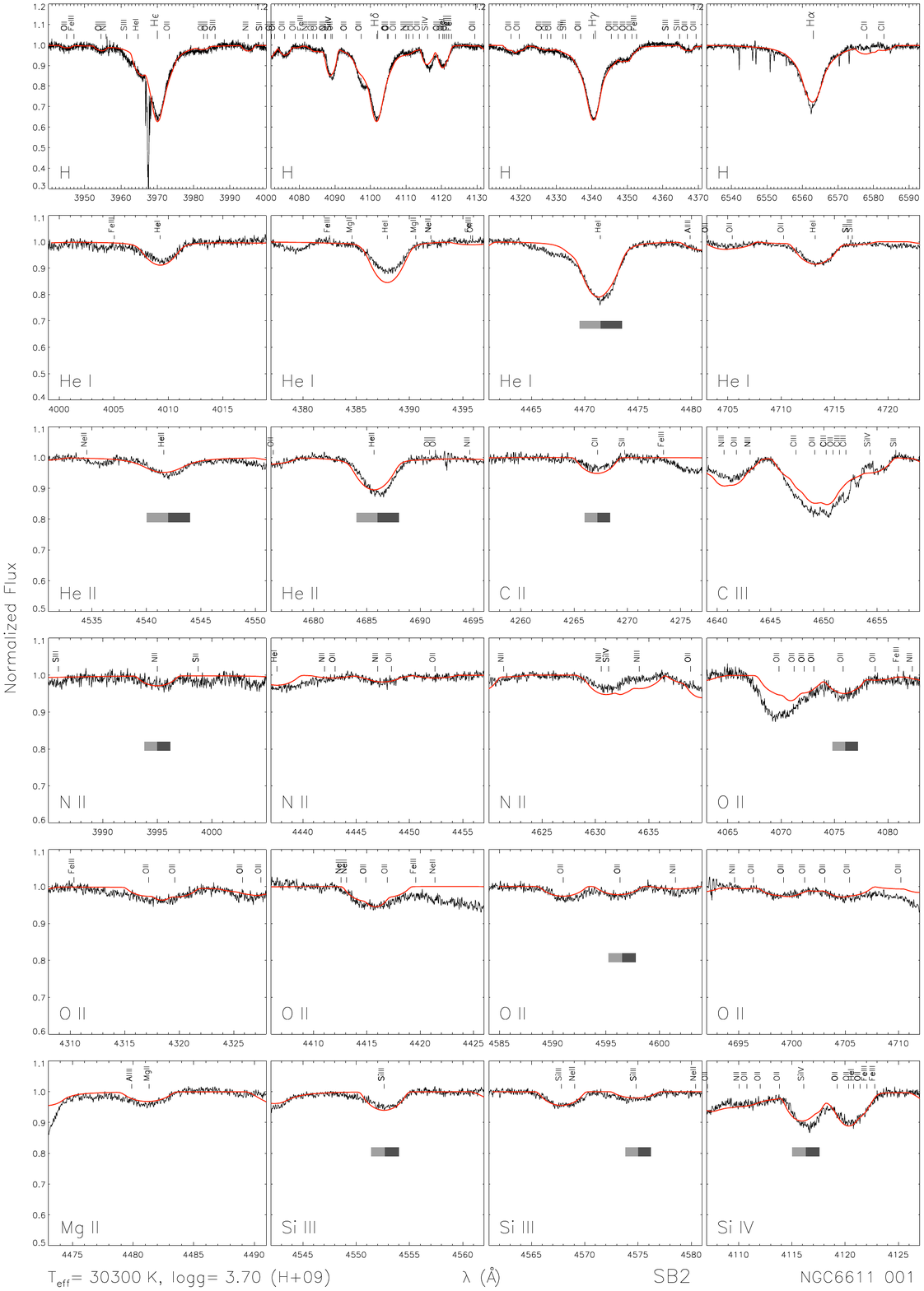}}
\vspace{-.5cm}
\caption[]{Same as Fig.~\ref{fit6611_006}, but for a FEROS observation of the star NGC\,6611-001. This is an example of a model fit to a double-lined binary (SB2) system, thus treating the object as a single star. The grey-shaded boxes guide the eye to recognise the binarity signatures in the most pronounced cases, \textit{i.e.},~the contributions of the two stars to the asymmetric line profiles. The neglect of the binary nature \citep[see also][]{Sanaetal09,Lefevreetal09} results in the analysis yielding incorrect atmospheric parameters and elemental abundances. Some deficits in the modelling,such as for the \ion{O}{ii} features around 4070\,{\AA}, stem from missing lines in the spectrum synthesis (see Fig.~\ref{fit6611_006}).}
\label{fit6611_001}
\end{figure*}

\begin{figure*}
\centering
\resizebox{0.89\hsize}{!}{\includegraphics[angle=0]{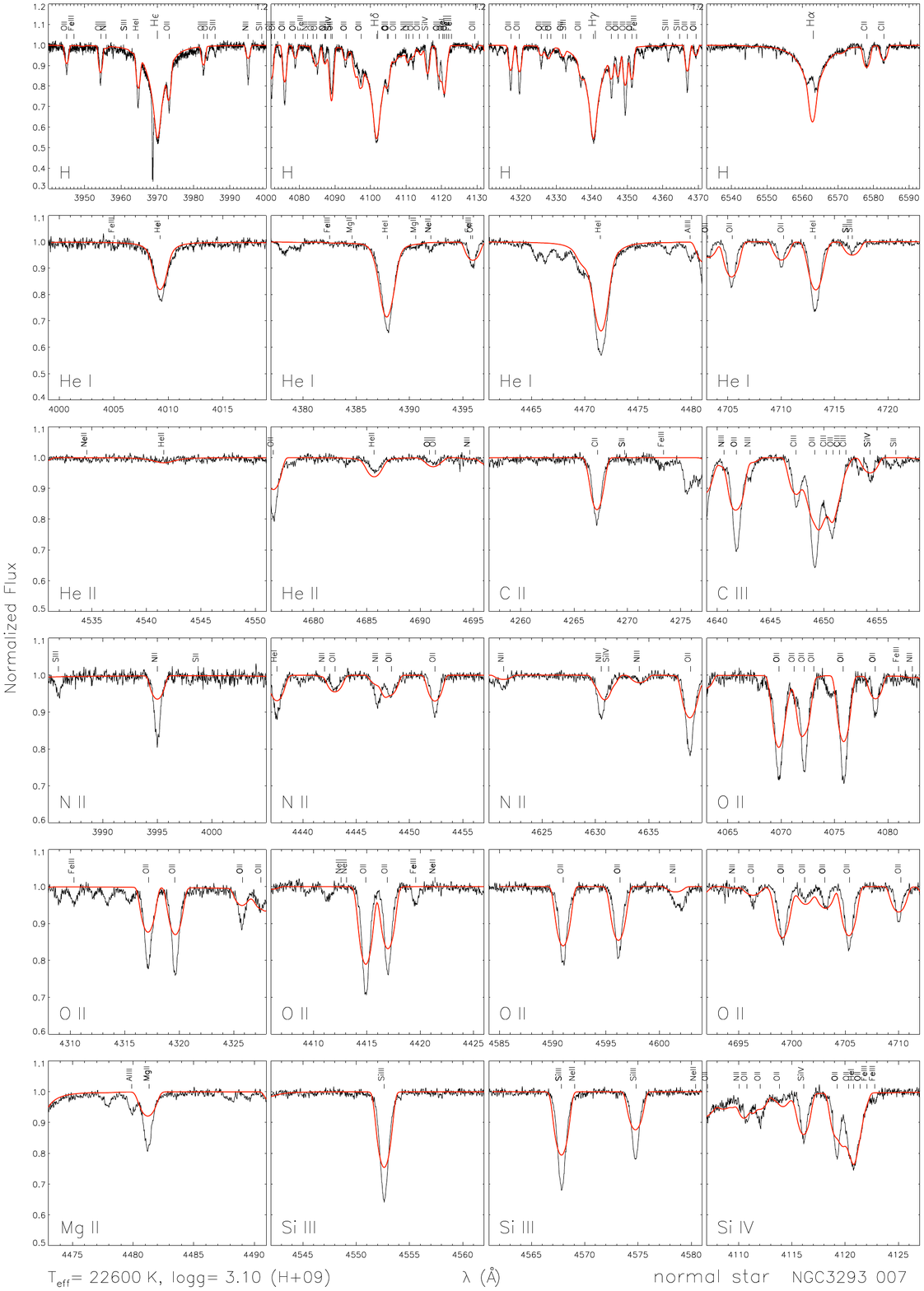}}
\vspace{-.5cm}
\caption[]{Same as Fig.~\ref{fit6611_006}, but for a FEROS observation of the star NGC\,3293-007. This is an example of a model fit to a nitrogen-poor evolved star \citep{Hunteretal09}. Except for the hydrogen Balmer lines (H$\alpha$ is filled by wind emission) the fit to all lines, and in particular to the \ion{N}{ii} lines, is poor. This indicates a poor choice for the stellar parameters, implying that the derived abundances are subject to systematic bias.}
\label{fit3293_007}
\end{figure*}

\begin{figure*}
\centering
\resizebox{0.89\hsize}{!}{\includegraphics[angle=0]{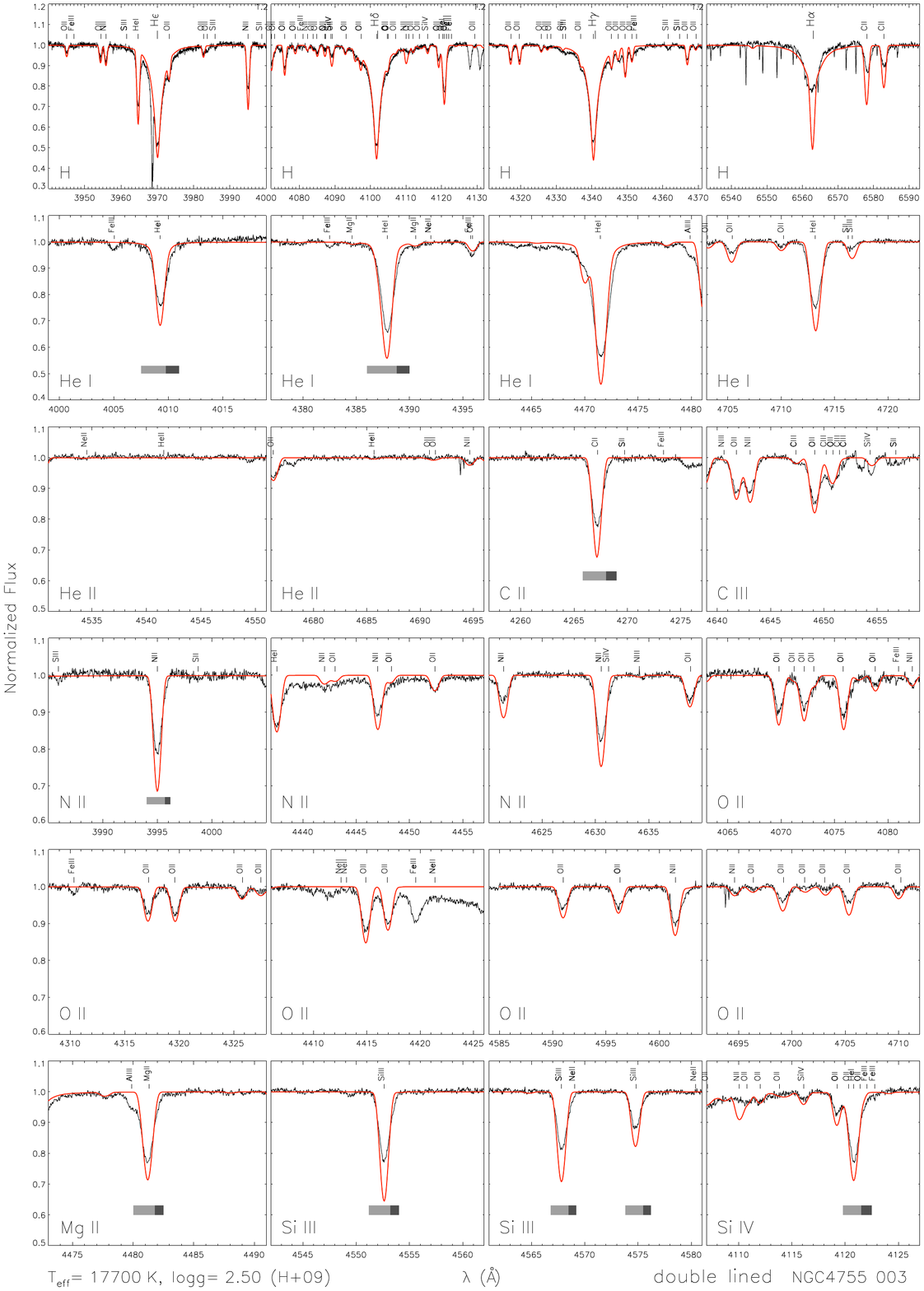}}
\vspace{-.5cm}
\caption[]{Same as Fig.~\ref{fit6611_006}, but for a FEROS observation of the star NGC\,4755-003. This is an example of a model fit to a nitrogen-rich evolved star \citep{Hunteretal09}. The fit to the \ion{He}{i} and most metal lines (including \ion{N}{ii}) is poor, indicating a poor choice for the stellar parameters. The derived elemental abundances are thus systematically biased. Moreover, close inspection finds the star to be double-lined. The grey-shaded boxes guide the eye to recognise the double-lined signatures in the most pronounced cases, \textit{i.e.},~the contributions of the two stars to the asymmetric line profiles.}
\label{fit4755_003}
\end{figure*}

We discuss six prototype examples representative of the four object classes identified in our investigations in Sect.~\ref{assessment} in the following. These comprise normal stars (on the MS and beyond), double-lined objects (SB2 candidates, visual or apparent binaries), Be stars, and a chemically peculiar object. The focus of the discussion, based on Figs.~\ref{fit6611_006} to \ref{fit4755_003}, is on the quality of the match of the synthetic with the observed spectra, in particular, with regard to the stellar parameter indicators -- the H and He lines, and metal ionisation equilibria. The individual panels of the figures are centred on the diverse diagnostic lines from different chemical species, sorted according to increasing atomic weight, and within one element by increasing ionisation degree and wavelength. Note that several -- mostly weaker -- metal lines are not implemented in the models, with the intention to reproduce only the FLAMES Survey~results. For the assessment of the entire sub-sample of 31 stars drawn from the \citetalias{Hunteretal09} work see Table~\ref{indicators}.
\begin{description}
\item[\sf NGC\,6611-006.] Figure~\ref{fit6611_006} shows an example for one of the best matches between theory and observation found within our re-investigation of the sub-sample of stars drawn from the \citetalias{Hunteretal09} work, for a single, normal star. Good agreement is found for the hydrogen line profiles, except for the narrow central H$\alpha$ emission that is of nebular origin. A good match is also obtained for the \ion{He}{i} and most metal lines, while a reasonable match is found for the \ion{He}{ii} and \ion{C}{iii} lines, and for \ion{Si}{iv}\,$\lambda$4654\,{\AA} (not considered by \citetalias{Hunteretal09}). Only \ion{C}{ii}\,$\lambda$4267\,{\AA} and \ion{C}{ii}\,$\lambda\lambda$6578/82\,{\AA} in the H$\alpha$ wing do not fit, with the predicted lines being too strong by a factor of $\sim$2 and a factor of several, respectively. The ionisation balance of \ion{C}{ii/iii} is thus not established, suggesting that an overall improved fit could be achieved by a revision of the atmospheric parameters (requiring also an adjustment of chemical abundances).
\item[\sf NGC\,2004-053.] This star is apparently the slowest rotator and yet one of the most nitrogen-rich stars in the LMC sample, making it a role-model for one of the star classes that cannot be explained on the basis of rotational mixing alone (\citetalias{Hunteretal09}). The broad central and symmetric H$\alpha$ emission in Fig.~\ref{fit2004_053} identifies the object as a Be star seen pole-on \citep[feature apparently overlooked by \citetalias{Hunteretal09}, but correctly identified by][]{Evansetal06}. Despite the sharp-lined character of the spectrum, this is, in fact, a fast-rotating star. The spectrum is dominated by light from the hot stellar pole, and smaller contributions from the gravity-darkened stellar equator regions (further diminished because of limb darkening) and the disc. The question is to which extent the non-uniform surface temperature and density, and the emission from the disc affect the continuum and line spectra. Despite the overall rather good match of model spectrum and observations (again except for \ion{C}{ii}~$\lambda$4267\,{\AA} and \ion{C}{ii}\,$\lambda\lambda$6578/82\,{\AA}, and a consistent failure to reproduce the depths of most metal lines) it is at present not possible to judge how realistic the derived parameters and abundances really are. Because of this potential systematic bias they should be treated with caution in any further interpretation. For the moment, the star should be excluded from further interpretation, like other Be-stars (see Sect.~\ref{assessment}). However, on observational grounds alone a true nature as a nitrogen-rich {\em fast} rotator is indicated.
\item[\sf NGC\,3293-034.] Despite being a good match for Balmer lines, \ion{He}{ii}\,$\lambda$4686\,{\AA} and many metal lines (Fig.~\ref{fit3293_034}) the stellar parameters and abundances derived by \citetalias{Hunteretal09} are not adequate. The reason for this is the mismatch of model and observation for the \ion{He}{i} lines. These indicate that NGC\,3293-034 is a He-strong star \citep[as identified by][]{Evansetal05}. Note also that none of the available ionisation equilibria (\ion{He}{i/ii}, \ion{C}{ii/iii}, \ion{Si}{iii/iv}) is matched in a satisfactory way. The higher molecular weight and lower opacity of He (with respect to H) in combination with a pronounced overabundance of the element change the atmospheric temperature and density structure, which requires dedicated model calculations for proper consideration beyond the pre-computed grids employed within the FLAMES Survey. In consequence, this star has to be excluded from further interpretation to avoid biased conclusions.
\item[\sf NGC\,6611-001.] The spectrum synthesis provides an overall good match to the observed spectrum, see Fig.~\ref{fit6611_001}, except for some \ion{He}{i/ii} lines and the unsatisfactory fit of the \ion{C}{iii}-dominated complex around 4650\,{\AA}. However, close inspection of isolated strong metal features like \ion{C}{ii}~$\lambda$4267\,{\AA}, \ion{O}{ii}~$\lambda$4076\,{\AA}, or \ion{Si}{iii}~$\lambda\lambda$4552, 4574\,{\AA} reveals line-profile asymmetries. Such asymmetries become apparent in particular in visual inspection when using line-profile fitting techniques, but are easily overlooked when employing an equivalent-width approach (like by \citetalias{Hunteretal09}) for the quantitative analysis. Line-profile asymmetries are indicators for a possible SB2 nature of the star, but require confirmation by time-series spectroscopy. Indeed, \citet{Sanaetal09} can clearly resolve double lines for NGC\,6611-001 for some epochs in their spectra \citep[note that the star is also known to be an eclipsing binary,][]{Lefevreetal09}. Consequently, the spectrum, while dominated by the light of the primary, is compromised by second light, and the analysis is therefore biased. To avoid misleading conclusions, the star has to be excluded from further interpretation.
\item[\sf NGC\,3293-007.] This is an example of an object evolved off the MS, identified as nitrogen-poor by \citetalias{Hunteretal09}. The comparison of observed and model spectra in Fig.~\ref{fit3293_007} shows an overall poor match, with the model largely under-predicting the depths of nearly all lines. In particular, the synthetic \ion{N}{ii} lines are too weak by a factor 2-3 in equivalent width, indicating that the finding of nitrogen-poorness is spurious. Moreover, the ionisation equilibria do not match, indicating poorly constrained stellar parameters and therefore, erroneous abundances in general. An independent comparison with non-LTE line-blanketed hydrodynamic models computed with {\sc Fastwind} and {\sc Cmfgen} \citep{HiMi98} by M.~Urbaneja (priv.~comm.) confirms a poor match of observation and models for the published parameters and abundances of \citetalias{Hunteretal09}. This is in line with the findings of \citet{MaPa13} that the abundances for Galactic OB stars determined by \citetalias{Hunteretal09}, on the one hand, and those determined by other codes differ \citep[despite some minor differences that also exist between the {\sc Ads}, {\sc Fastwind}, and {\sc Cmfgen} models]{Martinsetal08,Przy2011,NiSD11,FiPr12,NiPr12,Bouretetal12}. A $T_\text{eff}$ higher by 3000 to 4000\,K would seem more appropriate. The object should be excluded from further~interpretation.
\item[\sf NGC\,4755-003.] This is an example of an evolved nitrogen-rich star of \citetalias{Hunteretal09}. As in the case of the previous example, the match between observation and model is overall poor, see Fig.~\ref{fit4755_003}. This time, the model over-predicts the depths of almost all lines, including the \ion{N}{ii} features. Again, poorly constrained stellar parameters seem to be a major factor for the mismatch. The object should be excluded from further interpretation. A second reason for exclusion is found from closer inspection of the profiles of the stronger lines. They show the presence of a second absorption component in the red wings -- this object is double-lined. It appears that the second object is of similar spectral type but less luminous, probably a main-sequence star.
\end{description}

%: fig 18

%: fig 19

%: fig 20

%: fig 21

%: fig 22

%: fig 23
\end{appendix}

\end{document}